\documentclass[journal=aamick,manuscript=article]{achemso}
\usepackage{graphicx,subcaption,amsmath,gensymb,esvect,xcolor,xr}

\setkeys{acs}{maxauthors=30, etalmode=truncate}

   \makeatletter
\newcommand*{\addFileDependency}[1]{
  \typeout{(#1)}
  \@addtofilelist{#1}
  \IfFileExists{#1}{}{\typeout{No file #1.}}
}
\makeatother

\newcommand*{\myexternaldocument}[1]{%
    \externaldocument{#1}%
    \addFileDependency{#1.tex}%
    \addFileDependency{#1.aux}%
} 

\myexternaldocument{supp}

\usepackage[version=3]{mhchem}

\newenvironment{competing interests}
{
  \par\vspace{\baselineskip} \noindent
  \begin{Large}\textbf{Competing Interests}\end{Large} 
  \par \noindent\ignorespaces
}

\newenvironment{data availability}
{
  \par\vspace{\baselineskip}\noindent
  \begin{Large}\textbf{Data Availability}\end{Large}
  \par \noindent\ignorespaces
}

\newenvironment{author contribution}
{
  \par\vspace{\baselineskip}\noindent
  \begin{Large}{\textbf{Author Contribution}} \end{Large}
  \par \noindent\ignorespaces
}

\author{Shahid Sattar}
\email{shahid.sattar@ltu.se}
\affiliation{Applied Physics, Division of Materials Science, Department of Engineering Sciences and Mathematics,
Lule\aa\,University of Technology, Lule\aa\, SE-97187, Sweden}
\phone{+46 (0)920 491930}
\author{J. Andreas Larsson}
\email{andreas.1.larsson@ltu.se}

\title{Tunable Electronic Properties and Large Rashba Splittings Found in Few-Layer Bi$_2$Se$_3$/PtSe$_2$ Van der Waals Heterostructures}


\begin{document}


\begin{abstract}
We use first-principles calculations to show that van der Waals (vdW) heterostructures consisting of few-layer Bi$_2$Se$_3$ and PtSe$_2$ exhibit electronic and spintronics properties that can be tuned by varying the constituent layers. Type-II band alignment with layer-tunable band gaps and type-III band alignment with spin-splittings have been found. Most noticeably, we reveal the coexistence of Rashba-type spin-splittings (with large $\alpha_{\rm R}$ parameters) in both the conduction and valence band stemming from few-layer Bi$_2$Se$_3$ and PtSe$_2$, respectively, which has been confirmed by spin-texture plots. We discuss the role of inversion symmetry breaking, changes in orbital hybridization and spin-orbit coupling in altering electronic dispersion near the Fermi level. Since low-temperature growth mechanisms are available for both materials, we believe that few-layer Bi$_2$Se$_3$/PtSe$_2$ vdW heterostructures are feasible to realize experimentally, offering great potential for electronic and spintronics applications.      
\end{abstract}

\section{Introduction}
The rise of graphene undoubtedly served as a paradigm shift towards two-dimensional (2D) materials providing a versatile platform for future technological revolution \cite{c1,c2,c3}. Layer-by-layer assembly of transition metal dichalcogenides (TMDCs) soon after also enabled intelligent design possibilities for van der Waals (vdW) heterostructures of all dimensions revealing numerous exotic phenomena \cite{mos21,mos22,mos23,vdw1,vdw2,vdw3}.

Amongst a plethora of known 2D systems, few-layer PtSe$_2$ recently created substantial scientific interest due to the report of a range of properties, such as a layer-dependent band gap \cite{bandgap1}, high room-temperature electron mobility \cite{mobi1}, large stretchability \cite{str1}, which could be used in countless applications demonstrated for photocatalysis \cite{cat1} and optoelectronics \cite{opto1}. Different growth mechanisms have been adopted for high-quality PtSe$_2$ synthesis on various substrates including Pt(111) \cite{pt111}, silicon \cite{si}, sapphire \cite{saphire} and bilayer graphene/6H-SiC (0001) \cite{graphene} to list a few. Recently, complementary metal-oxide-semiconductor compatible large-scale fabrication of a trilayer PtSe$_2$ MOSFET has been demonstrated with a current ON/OFF ratio approaching 1600 at 80 K hinting at improvements in 2D nanoelectronics \cite{mosfet}. Moreover, monolayer PtSe$_2$ has been shown to host helical spin-texture and show local dipole induced Rashba effect with spin-layer locking which is advantageous for electrically controllable spintronics devices \cite{spinptse2}. Heterostructures consisting of vertically stacked PtSe$_2$/MoSe$_2$ show type II band alignment and interface states originating from the strong-weak interlayer coupling of the constituent systems \cite{ptvdw}.

Three-dimensional (3D) topological insulator Bi$_2$Se$_3$ is also a promising material for spintronic applications owing to room-temperature spin-polarized surface currents \cite{li2014electrical,dankert2015room}, efficient charge-to-spin conversion via doping \cite{dankert2018origin}, giant spin pumping and inverse spin Hall effects through ferromagnetic contacts \cite{jamali2015giant}. Moreover, few-layer Bi$_2$Se$_3$ has been shown to demonstrate interesting properties, e.g., thickness-modulated semiconducting behavior \cite{bise1} in comparison to its topologically insulating (TI) bulk counterpart \cite{bise2,bise3}. Several novel functionalities have been enabled using few-layer Bi$_2$Se$_3$ (e.g., coexistence of topological order and superconductivity \cite{sup-top}, hedgehog spin texture and Berry phase tuning \cite{berry}, thermoelectrics \cite{thermo}, and ultrafast carrier dynamics, \cite{ultfast}). Furthermore, few-layer Bi$_2$Se$_3$ also shows sizable Rashba-type spin-splittings due to substrate-induced structural inversion asymmetry \cite{bise1,bise-r,alpha1,bise-r2}. 

As both few-layer Bi$_2$Se$_3$ and PtSe$_2$ demonstrate layer-dependent properties and have several distinctive features, it is therefore of great interest to theoretically study their vdW heterostructures (by varying the number of constituent layers) and search for possible synergy effects that could be utilized. Interestingly, both materials have low temperature growth mechanisms (up to 450 \degree C), thus, it is likely that such heterostructures can be accomplished experimentally. Motivated by this, we employ first-principles calculations and unfold tunable and sizeable type-II and type-III band alignments as well as several different spintronics features of few-layer Bi$_2$Se$_3$/PtSe$_2$ vdW heterostructures. We reveal the coexistence of Rashba-type spin-splittings in the conduction(valence) band originating from few-layer Bi$_2$Se$_3$(PtSe$_2$) due to inversion symmetry breaking and structural asymmetry. Our findings provide a promising pathway to manipulate the charge and spin degrees of freedom using carefully designed vdW heterostructures for the next-generation nanoscale electronic and spintronics devices. 

\begin{figure}[!t]
\includegraphics[width=1.0\textwidth]{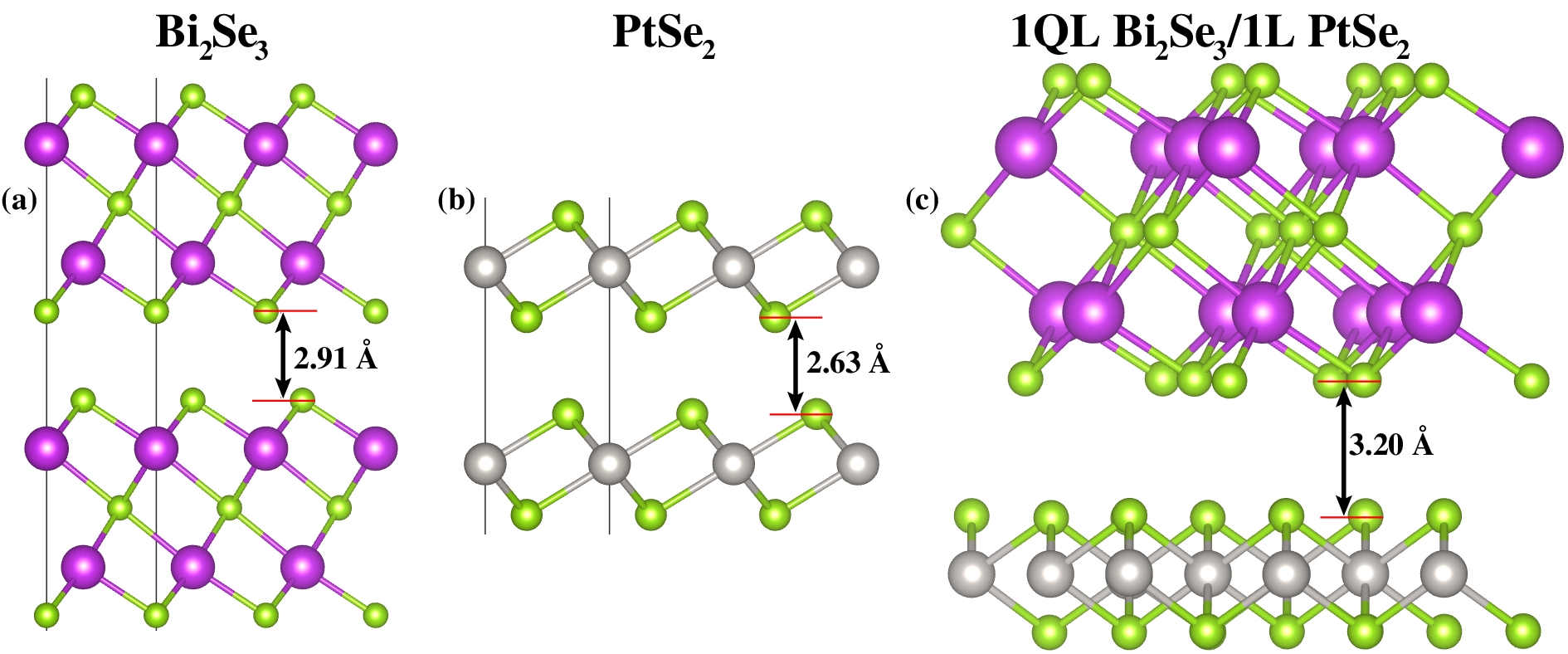}
\caption{Side views of (a) few-layer Bi$_2$Se$_3$, and (b) PtSe$_2$ (black lines depict unit cell) alongside (c) 1QL Bi$_2$Se$_3$/1L PtSe$_2$ as a representative vdW heterostructures considered in this study. The interlayer distance is given for each case. Pt, Se, and Bi atoms are shown in grey, green, and purple, respectively.}
\label{fig:fig1}
\end{figure}

\section{Computational method}

We have performed density functional theory (DFT) calculations using the projector augmented wave method\,\cite{paw1,paw2} as implemented in the Vienna Ab-initio Simulation Package\,\cite{vasp}. For the exchange-correlation potential, we have used the non-local optB86b-vdW density functional \cite{optb86,optb86-2} to account for the van der Waals interactions between the layers as successfully employed before for similar systems \cite{peng2016electronic,bi2te3}. The plane-wave cutoff energy was set to a sufficiently large value of 420 eV. A gamma-centered Monkhorst-Pack $4\times 4\times 1$ k-mesh was used for the structural relaxation whereas for the band structure calculations, the Brillouin zone integration was performed using a dense $7\times7\times 1$ k-mesh. Moreover, to compute 2D spin-textures, we set up a 2D k-mesh ($k_x\times k_y:15\times15$) centered at the gamma-point ($k_z=0$). For the iterative solution of the Kohn-Sham equations, we ensured the total energy to converge until the change is below $10^{-6}$ eV and residual forces on the atoms to decline to less than $10^{-3}$ eV/\AA. Since our systems contain heavy elements, the effects of spin-orbit coupling (SOC) were taken into account in the band structure and density of states (DOS) calculations. The heterostructures were modeled using a 15\,\AA\,thick vacuum layer in the out-of-plane direction to avoid periodic images interactions. Finally, the PyProcar python library \cite{pyprocar} and Matplotlib graphics package \cite{matplotlib} were used for pre- and post-processing of the data and plotting.

\section{Results and Discussion}

We have used DFT to simulate single- and few-layer Bi$_2$Se$_3$ and PtSe$_2$ and their heterostructures, as shown in Figure \ref{fig:fig1}(a-c). Each Bi$_2$Se$_3$ slab consists of five atomic sheets (termed a quintuple layer (QL)) having Se-Bi-Se-Bi-Se atoms held together by covalent bonding, whereas a Pt-atom is covalently sandwiched between two Se-atom-layers in a PtSe$_2$ slab with 1T-phase trigonal geometry. The optimized lattice parameters for 1QL(2QL) Bi$_2$Se$_3$ and 1L(2L) PtSe$_2$ are 4.15\,\AA\,(4.14\,\AA) and  3.71\,\AA\,(3.74\,\AA\,), respectively, in close agreement with reported values. Few-layer Bi$_2$Se$_3$ is separated by 2.91\,\AA\,whereas Pt chalcogenides are known to experience strong interlayer interactions as evident from an interlayer distance of 2.63\,\AA. In order to achieve minimal lattice mismatch in forming vdW heterostructures, a $\sqrt{7}\times\sqrt{7}\times1R19.1\degree$ supercell of Bi$_2$Se$_3$ is matched to a $3\times3\times1$ supercell of PtSe$_2$ with a lattice parameter of 11.05\,\AA\,for the combined supercell resulting in less than 1\% lattice mismatch. Similarly for other heterostructures, average lattice parameters are adopted for vdW heterostructures in subsequent calculations. Owing to the incommensurate nature of the 2D materials involved in our study, a Moir\'{e} superlattice of periodicity 1.1 nm is therefore expected to be obtained in experiments, similar to the case of PtSe$_2$/MoSe$_2$ heterostructures \cite{ptvdw}. The lowest-energy structure for a representative 1QL Bi$_2$Se$_3$/1L PtSe$_2$ heteroestructure is given in Figure \ref{fig:fig1}(c), for which different lateral stackings were carefully inspected before arriving at this configuration. We note that interlayer interactions between PtSe$_2$ and Bi$_2$Se$_3$ are homogeneous meaning Bi or Se atoms do not prefer particular sites on PtSe$_2$. For six different lateral stacking configurations, shown in supplementary Figure \ref{fig:figs1}(a-f), the energy differences fall in the of range 0 to 10 meV from which the minimum energy configuration is adopted for calculations.

\begin{figure}[!t]
\includegraphics[width=1.0\textwidth]{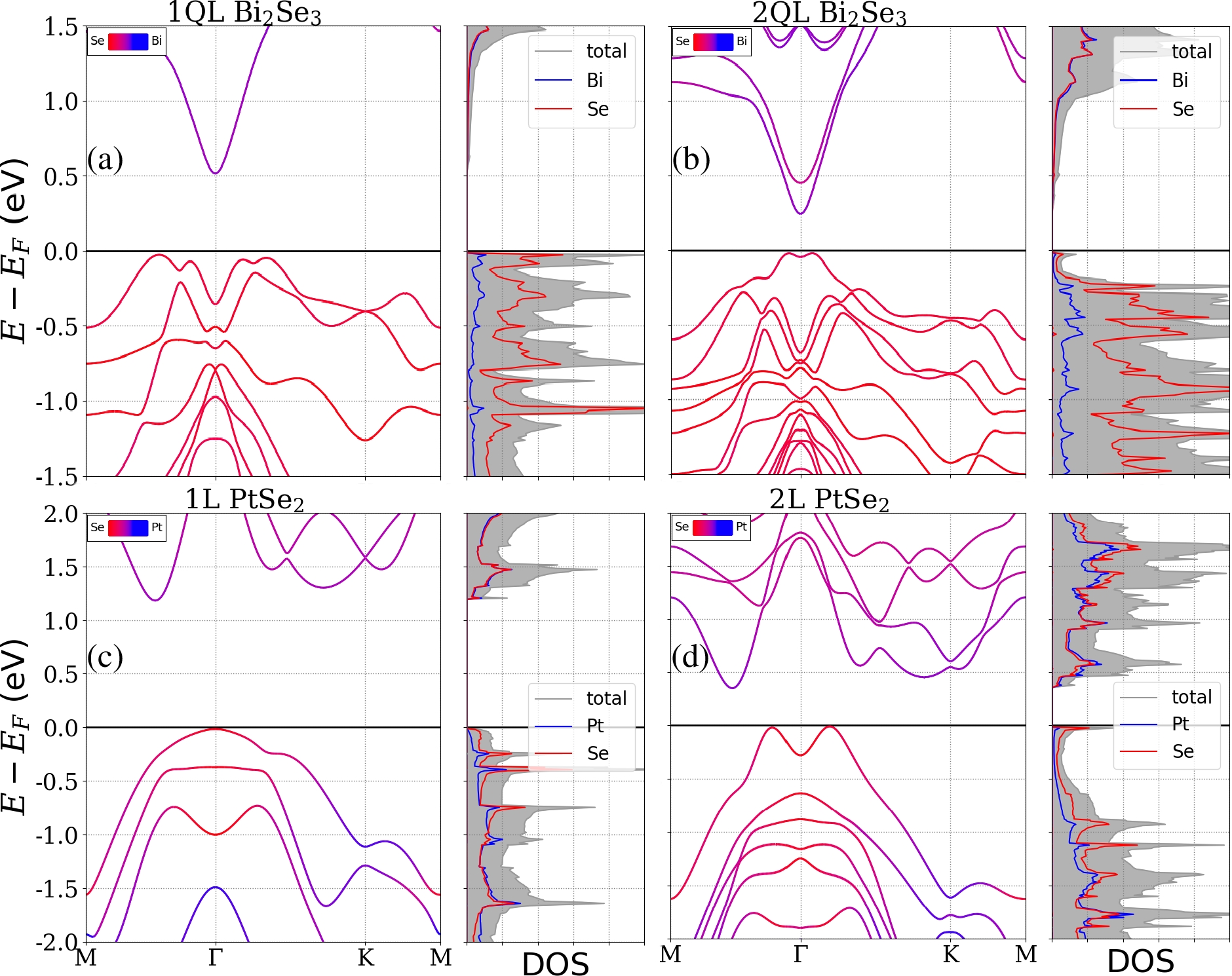}
\caption{Band structures and total/atom projected density of states (DOS) of (a,b) few-layer Bi$_2$Se$_3$, and (c,d) few-layer PtSe$_2$, respectively. Spin-orbit coupling (SOC) is taken into account in all cases.}
\label{fig:fig2}
\end{figure}

We first determine the electronic properties of pristine few-layer Bi$_2$Se$_3$ and PtSe$_2$ (see Figure \ref{fig:fig2}). A layer-dependent band gap with decreasing magnitude is observed in both cases. For 1QL(2QL) Bi$_2$Se$_3$, incorporating the effect of SOC, an indirect band gap of 0.55 eV(0.26 eV) is observed as shown in Figure \ref{fig:fig2}(a,b), respectively. The conduction band (CB) minimum for 1QL is located at the high-symmetry $\Gamma$-point whereas the valence band (VB) maximum lies along the $\Gamma$-M direction. Looking at the atom projected DOS (PDOS), the VB consists mainly of Se$(p$) states and small contribution from Bi$(d)$ states whereas the CB contains evenly mixed Se$(p)$-Bi$(d)$ states. For 1L(2L) PtSe$_2$, we observe a comparatively larger indirect band gap of 1.20 eV(0.36 eV) with the CB minimum lying along the $\Gamma$-M direction and the VB maximum is located at the high-symmetry $\Gamma$-point for 1L (see Figure \ref{fig:fig2}(c,d)). For 1L PtSe$_2$, the VB comprises also of Se$(p$) states and minuscule Pt$(d)$ contribution present near the Fermi level that gradually escalates in magnitude in moving away from it. The CB of few-layer PtSe$_2$ however also involves equally mixed Se$(p)$-Pt$(d)$ states much alike few-layer Bi$_2$Se$_3$. The reason for the differences between 1(Q)L and 2(Q)L is that although the strong interlayer binding is of van der Waals type, the resulting physisorption leads to induced moments and Pauli repulsion of the electron density (contracted in the layers) that effects the properties since the layers are squeezed together compared to a monolayer \cite{and1}. Unlike other TMDCs such as MoS$_2$, layer-tunable band gaps and mixed atomic hybridization in both few-layer Bi$_2$Se$_3$ and PtSe$_2$ influences the observed novel features described above, as will be discussed in the following section.  

We have considered four different heterostructures combining few-layer Bi$_2$Se$_3$ and PtSe$_2$, denoted nQL Bi$_2$Se$_3$/nL PtSe$_2$ (see Table \ref{table:tabl1}), for which we report the structural and energetic properties. The binding energy per PtSe$_2$ is calculated through equation (\ref{equation:equation1}),
\begin{equation}
E_b=(E[{\rm nQL\,Bi_2Se_3/\rm nL\,PtSe_2}]-E[{\rm nQL\, Bi_2Se_3}]-E[{\rm nL\,PtSe_2}])/3, \label{equation:equation1}
\end{equation}
where $E[{\rm nQL\,Bi_2Se_3/\rm nL\,PtSe_2}]$ is the total energy of the few-layer vdW heterostructure, $E[{\rm nQL\,Bi_2Se_3}]$ is the total energy of detached nQL Bi$_2$Se$_3$, and $E[{\rm nL\, PtSe_2}]$ is the total energy of detached nL PtSe$_2$ considered in the heterostructures, where n$=1,2$. We have found the layers of Bi$_2$Se$_3$ and PtSe$_2$ to bind through physisorption, as expected, and these are thus vdW heterostructures. 

\begin{table}[ht]
\caption{Lattice parameter (a), interlayer distance ($d$), binding energy ($E_b$), band gap  ($E_g$), and band alignment (B$_{\rm Alignment}$) of few-layer Bi$_2$Se$_3$ and PtSe$_2$ vdW heterostructures.} 
\centering 
\begin{tabular}{c  c  c  c  c  c} 
\hline\hline 
Heterostructure &\,\, $a(\mathrm{\AA})$ &\,\, $d(\mathrm{\AA})$ &\,\, $E_b(eV)$ &\,\, $E_g(meV)$&\,\,B$_{\rm Alignment}$ \\ [0.5ex] 
\hline 
1QL Bi$_2$Se$_3$/1L PtSe$_2$ \,\,\,\,&\,\, 11.05 &\,\, 3.20 &\,\, -0.63 &\,\, 100&\,\,Type-II \\ 
2QL Bi$_2$Se$_3$/1L PtSe$_2$ \,\,\,\,&\,\, 11.04 &\,\, 3.17 &\,\, -0.63 &\,\, 50&\,\,Type-II \\
1QL Bi$_2$Se$_3$/2L PtSe$_2$ \,\,\,\,&\,\, 11.10 &\,\, 3.12 &\,\, -0.66 &\,\, $-$&\,\,Type-III \\
2QL Bi$_2$Se$_3$/2L PtSe$_2$ \,\,\,\,&\,\, 11.08 &\,\, 3.12 &\,\, -0.66 &\,\, $-$&\,\,Type-III \\ [1ex] 
\hline 
\end{tabular}
\label{table:tabl1} 
\end{table}

\begin{figure}[!htb]
\includegraphics[width=0.9\textwidth]{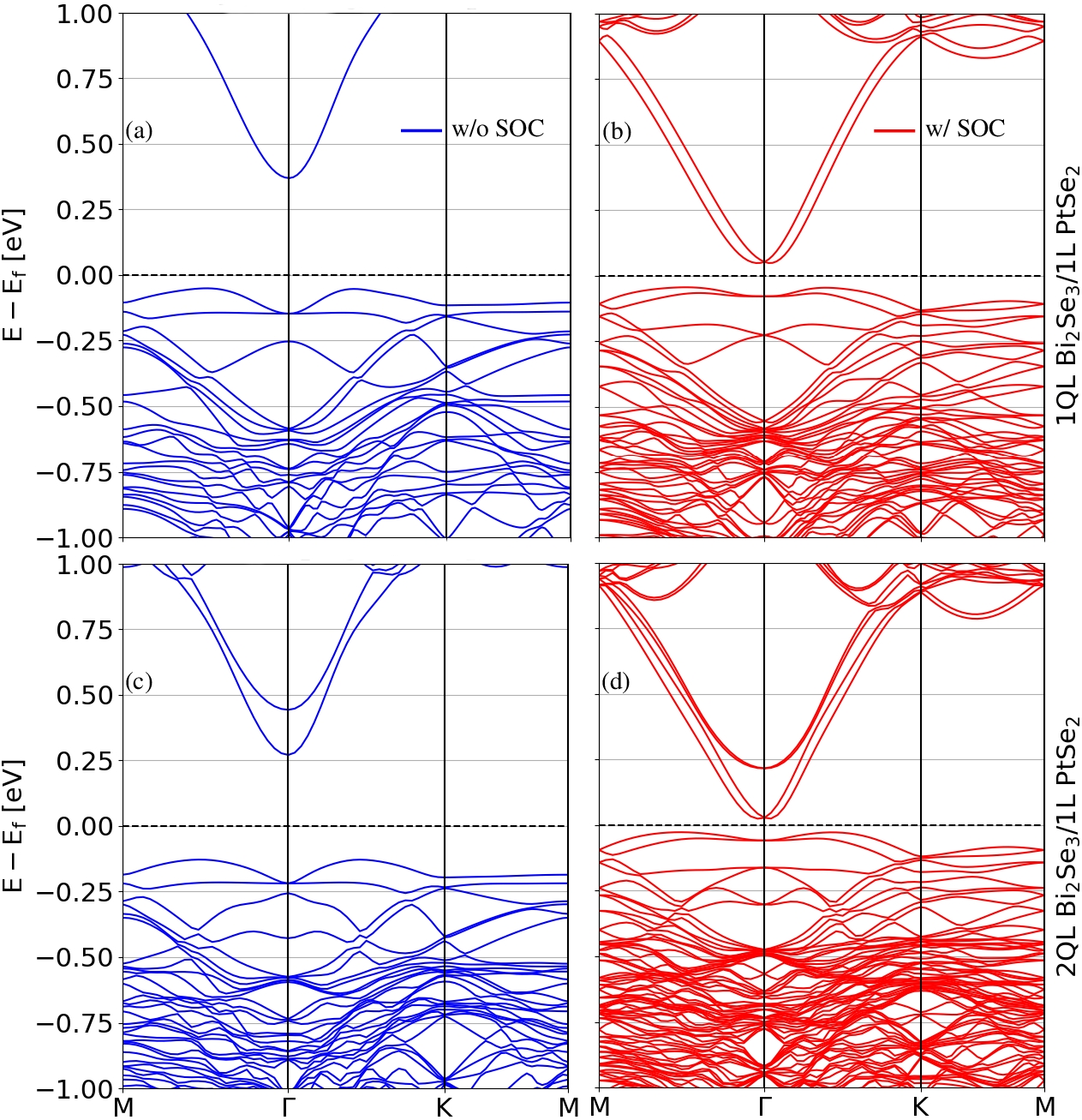}
\caption{(a-d) Band structures of 1QL(2QL) Bi$_2$Se$_3$/1L PtSe$_2$ vdW heterostructures (without SOC (left,blue) and with SOC (right,red)).}
\label{fig:fig3}
\end{figure}

Looking at the binding energy ($E_b$), it is clear that 1L PtSe$_2$ binds much stronger to 1QL Bi$_2$Se$_3$ (-0.63 eV/PtSe$_2$) in comparison to the recently studied case of 1L MoSe$_2$/1L PtSe$_2$ (-0.25 eV/PtSe$_2$) \cite{ptvdw}. Moreover, we observe that increasing the PtSe$_2$ thickness further enhances the interlayer coupling as $E_b$ increases to -0.66 eV/PtSe$_2$ in moving from 1L to 2L PtSe$_2$. The contraction in interlayer distance (3.20\,\AA\,to 3.12\,\AA) also validate this argument. On the other hand, adding a second QL of Bi$_2$Se$_3$ to the heterostructure shows negligible change in both binding energy and interlayer distance. It is worth mentioning that the interlayer distance in vdW heterostructures is always longer than pristine few-layer Bi$_2$Se$_3$ and PtSe$_2$ showing relatively weaker interaction than for the constituent systems. 

\begin{figure}[!htb]
\includegraphics[width=1.0\textwidth]{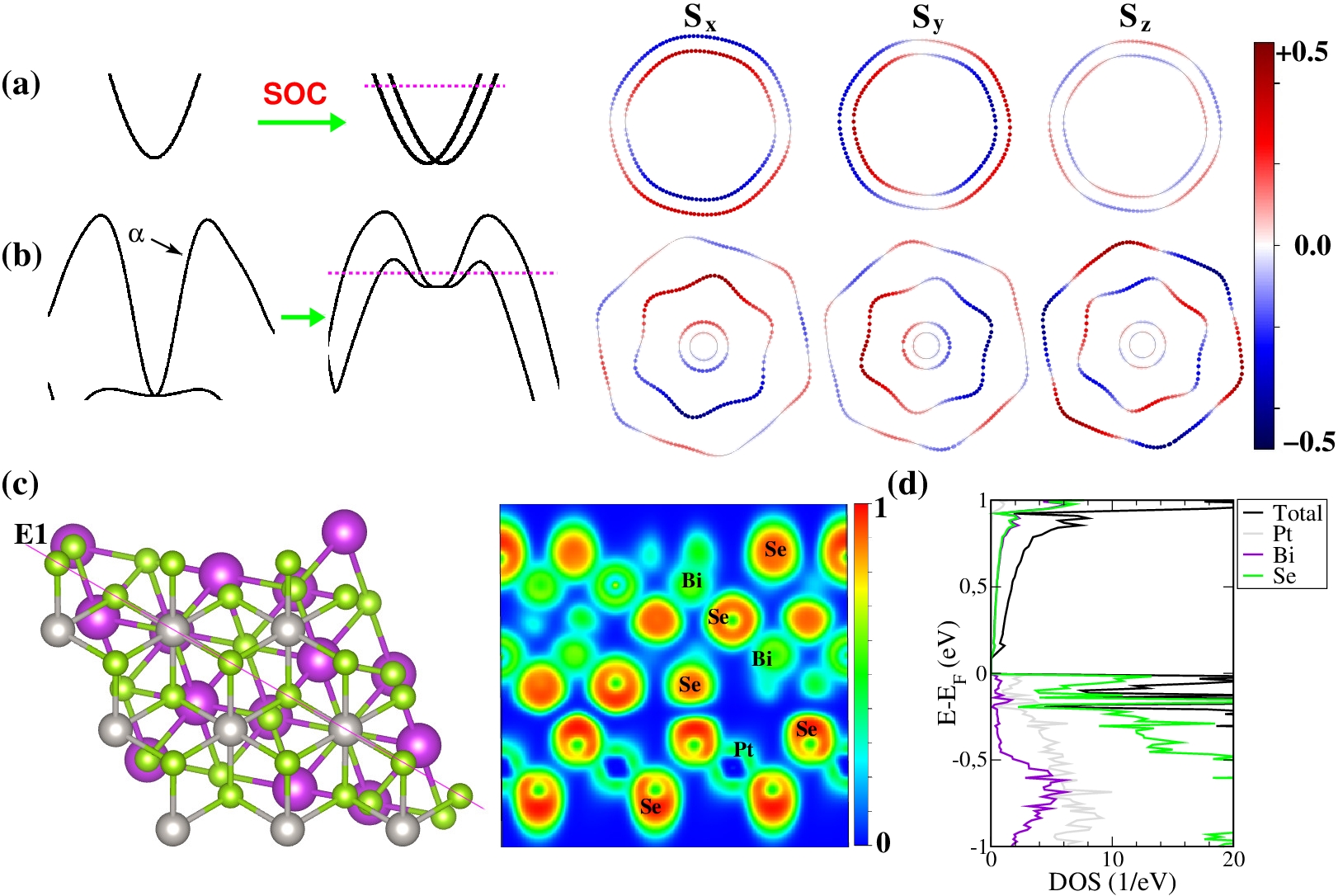}
\caption{For 1QL Bi$_2$Se$_3$/1L PtSe$_2$: (a,b) stylized CB and VB regions without and with SOC alongside fixed-energy (magenta dotted line: 0.4 eV for the CB and -0.080 eV for the VB) contour plots of spin-components $s_x$, $s_y$ and $s_z$. (c) 2D Electron localization function (ELF) plot corresponding to the purple line E1. (d) Total and atom projected DOS for the said vdW heterostructure.}
\label{fig:fig4}
\end{figure}

For the 1QL(2QL) Bi$_2$Se$_3$/1L PtSe$_2$ heterostructure, including the effect of SOC (see Figure \ref{fig:fig3}(b,d)), we found a type-II band alignment with layer-tunable indirect band gaps of 100 meV(50 meV), respectively. The CB minimum is present at the high symmetry $\Gamma$-point coming from few-layer Bi$_2$Se$_3$ and the VB is contributed by 1L PtSe$_2$ having its maximum along the $\Gamma$-M direction (see supplementary Figure \ref{fig:figs2}(a)). Comparing to the case without SOC (Figure \ref{fig:fig3}(a,c)), the SOC significantly affects the magnitude of the band gaps in the vdW heterostructures much alike pristine 1QL(2QL) Bi$_2$Se$_3$ (see supplementary Figure \ref{fig:figs3}(a-d)). For the 1QL(2QL) Bi$_2$Se$_3$/1L PtSe$_2$ heterostructures the energy gap shrinks with the type-II band alignment when going from 1QL to 2QL, which is in line with the change of the CB in pristine Bi$_2$Se$_3$ in 1QL and 2QL (c.f. Figure \ref{fig:fig2}(a,b)). In the context of theoretical predictions of strongly bound excitons in 1L PtSe$_2$ \cite{exciton} and the recent experimental observation of chiral surface excitons in a TI Bi$_2$Se$_3$ \cite{chiral}, our findings are highly intriguing because they hint at possible control of low-energy interlayer excitons by forming layer-controlled vdW heterostructures. Moreover, following the trend of the band gap values, it is anticipated that the band gap will eventually vanish at around 4QLs Bi$_2$Se$_3$/1L PtSe$_2$ vdW heterostructure. 

Interestingly, besides layer-tunable type-II band alignment, we also observe coexistence of Rashba-type spin-splittings (in the CB) and hedgehog-like band-splittings (in the VB) of 1QL(2QL) Bi$_2$Se$_3$/1L PtSe$_2$ vdW heterostructures. The crystal structures of few-layer Bi$_2$Se$_3$ have inversion symmetry (i.e., $(x,y,z) \longrightarrow (-x,-y,-z)$) which does not remain intact for vdW heterostructures with 1L PtSe$_2$. Applying SOC therefore lifts the spin-degeneracy and display Rashba-type spin-splittings in the CB originating from few-layer Bi$_2$Se$_3$. We note that these splittings resemble to those observed in graphene/TI vdW heterostructures resulting in gate-tunable spin-galvanic effects at room temperature \cite{khokhriakov2020gate}. To confirm this, we set up a $\Gamma$-centered 2D k-mesh along the xy-plane and plot the fixed-energy contours of spin components $s_x$, $s_y$ and $s_z$ as shown in Figure \ref{fig:fig4}(a) alongside the transformation of the CB with respect to the SOC effect. The typical Rashba-split bands along the momentum-axis with large(small) in-plane(out-of-plane) spin components, respectively, support our findings. Using the Rashba Hamiltonian for 2D-electron gas, 
\begin{equation}
    H_R=\pm \frac{\hbar^2k_{\parallel}^2}{2m^*}+\alpha_R\vv{\sigma} . (\vv{k}_{\parallel}\times \vv{z}),
    \label{equation:equation2}
\end{equation}
where $k_{\parallel}=(k_x,k_y,0)$ and $m^*$ being the in-plane momentum and effective mass of electron, $\alpha_R$ is the Rashba parameter, $\vv{\sigma}$ is the vector of Pauli matrices and $\vv{z}$ is the out-of-plane unit vector. Taking $E_R$ as the energy difference between the CB minimum and band crossing at the $\Gamma$-point and $k_0$ as the momentum offset, the Rashba parameter for a parabolic-dispersion is approximated by $\alpha_R=2E_R/k_0$, whereas $E_R=\hbar^2k_{0}^2/2m^*$ and $k_0=m^*\alpha_R/\hbar^2$. For 1QL(2QL) Bi$_2$Se$_3$/1L PtSe$_2$ vdW heterostructures, we obtain $E_R=4.8\,meV(4.0\, meV)$ and $k_0=0.002\,\mathrm{\AA}^{-1}(0.002\,\mathrm{\AA}^{-1})$ thus giving $\alpha_R=4.8\,eV\,\mathrm{\AA}(4.0\,eV\, \mathrm{\AA})$, respectively. These values for the Rashba parameter ($\alpha_R$) are amongst the highest found, compared to similar 2D vdW heterostructures (see supplementary Table \ref{table:tables1}). Moreover, a slight decrease in $\alpha_R$ moving from 1QL to 2QL Bi$_2$Se$_3$ suggests layer-tunable spintronics in these vdW heterostructures. Furthermore, another striking feature is the presence of a giant energy interval of 0.9 eV above the Fermi level and inside the band gap of 1L PtSe$_2$ without any non-Rashba states. This constitutes a complete 2D Rashba electron-gas in 1QL Bi$_2$Se$_3$/1L PtSe$_2$ vdW heterostructure further extending the limit from Ref. \cite{bise-r2}. Having consistent behavior for 2QL Bi$_2$Se$_3$/1L PtSe$_2$ with lesser magnitude (0.2 eV) of pure Rashba-split CB, our results point towards a promising route to design Datta Das spin field-effect transistor out of these few-layer vdW heterostructures. 

\begin{figure}[!htb]
\includegraphics[width=0.9\textwidth]{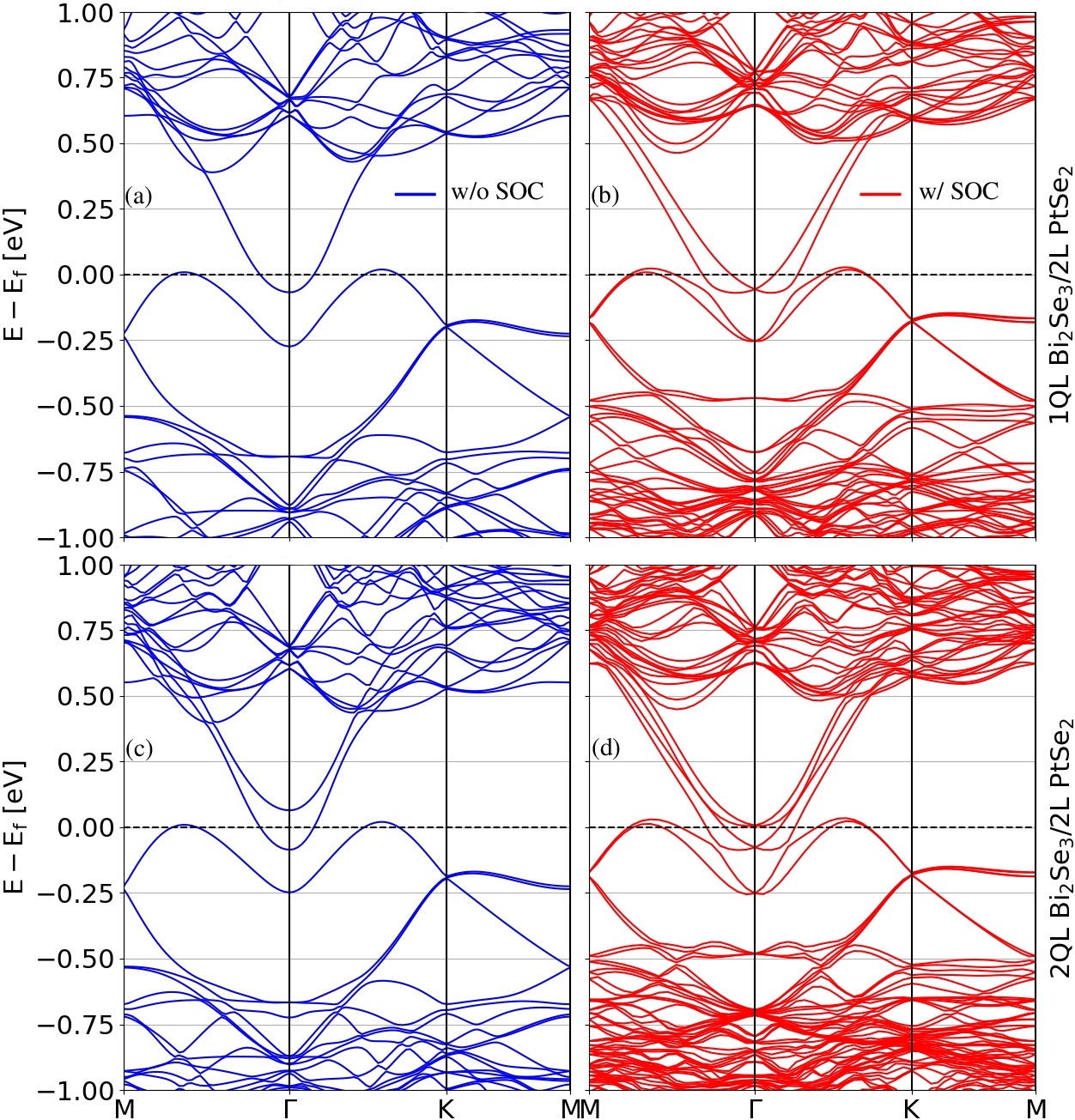}
\caption{(a-d) Band Structures of 1QL(2QL) Bi$_2$Se$_3$/2L PtSe$_2$ vdW heterostructures (without SOC (left,blue) and with SOC (right,red)).}
\label{fig:fig5}
\end{figure}

On the other end, looking at the topmost VB of the band structure coming from 1L PtSe$_2$, it is seen turning into a ``Mexican hat" shape (compare Figure \ref{fig:fig3}(a,c) to supplementary Figure \ref{fig:figs4}(c) for the band structure of pristine 1L PtSe$_2$). The upper VB of 1L PtSe$_2$ (called $\alpha$-band, see Figure \ref{fig:fig4}(b)) is mainly formed from the in-plane $p_{xy}$-orbitals of Se atoms. We have analyzed the interlayer interaction using the electron localization function (EFL), as can be seen in Figure \ref{fig:fig4}(c), and find Bi$_2$Se$_3$ to be bound with PtSe$_2$ by vdW physisorption. The electron density of a 1QL(2QL) Bi$_2$Se$_3$/1L PtSe$_2$ shrinks compared to a single layer due to the induced moments and Pauli repulsion as mentioned above and shown for graphene/graphite in Ref. \cite{and1}. This contraction can be seen comparing the interface with the outer surfaces of Bi$_2$Se$_3$ and PtSe$_2$ in Figure \ref{fig:fig4}(c), and leads to perturbations of the chemical bonds within the layers and results in changes to the band structure as seen in Figure \ref{fig:fig4}(b). Moreover, the total and atom projected DOS shown in Figure \ref{fig:fig4}(d) confirm the changes in orbital hybridization between Se$(p)$- and Bi/Pt$(d)$-states near the Fermi level compared to Figure \ref{fig:fig2}. Moreover, orbital-projected band structures of 1QL Bi$_2$Se$_3$/1L PtSe$_2$ vdW heterostructure, given in the supplementary Figure \ref{fig:figs2prime}(a,b), show much larger Se-orbital hybridization in the VB compared to the CB. In the 1T-phase, 1L PtSe$_2$ holds centrosymmetry with D$_{3d}$ point group which does not remain intact due to the different charge environment experienced by the top and bottom Se-layers. Applying SOC therefore lifts the spin-degeneracy and produce hedgehog-like band-splittings without any net spin polarization for which the underlying theoretical formalism is discussed in Ref. \cite{bsvsp} and resembles that of the recently realized experiment of Ref. \cite{tell-prl}. To make this point clear, we show the transformation of the VB due to SOC along fixed-energy contours of the spin components $s_x$, $s_y$ and $s_z$ in Figure \ref{fig:fig4}(b). One may also distinguish between Rashba-type spin-splittings of the CB and hedgehog-like band-splittings of the VB by looking at the fixed-energy contours of the spin-components in both Figure \ref{fig:fig4}(a) and \ref{fig:fig4}(b) right-side, respectively. 

We also considered the effect of increasing the PtSe$_2$ thickness, i.e., few-layer Bi$_2$Se$_3$/2L PtSe$_2$, for which the band structures show type-III band alignment as displayed in Figure \ref{fig:fig5}(a-d) and supplementary Figure \ref{fig:figs2}(b). Interestingly, both the CB and VB simultaneously show Rashba-type spin-splittings with each band being located on the separate constituent materials. The broken inversion symmetry is valid as for 1QL(2QL) Bi$_2$Se$_3$/2L PtSe$_2$ (i.e., top and bottom constituent layers experience different charge environment). Approximating the band dispersion around the $\Gamma$-point by Eq.\,\ref{equation:equation2}, Table \ref{table:table2} lists the corresponding parameters for all vdW heterostructures considered in this study. Most notably, at a momentum offset $k_0=0.033$\,\AA$^{-1}$ and giant Rashba spin-splitting energy of $275-278$ meV, the resultant $\alpha_R=16.66-16.84$ eV\,\AA\, for few-layer Bi$_2$Se$_3$/2L PtSe$_2$ are amongst the highest values reported to date (see supplementary Table \ref{table:tables1}). Moreover, the $\alpha_R$ values for the CB also show the formation of a Rashba electron-gas over a large energy interval. 

\begin{table}[ht]
\caption{Rashba spin-splitting parameters of few-layer Bi$_2$Se$_3$/PtSe$_2$ vdW heterostructures. $E_R$ is energy difference between the CB/VB minimum/maximum and band crossing at the $\Gamma$-point, $k_0$ is the momentum shift and $\alpha_R$ is the Rashba parameter.} 
\centering 
\begin{tabular}{c  c  c  c} 
\hline\hline 
Heterostructure & $E_R(\text{meV})$ & $k_0(\mathrm{\AA}^{-1})$ & $\alpha_R(\text{eV}\,\mathrm{\AA})$ \\ [0.5ex] 
\hline 
1QL Bi$_2$Se$_3$/1L PtSe$_2$ & 4.8 & 0.002 & 4.80  (CB) \\
2QL Bi$_2$Se$_3$/1L PtSe$_2$ & 4.0 & 0.002 & 4.00  (CB) \\
2QL Bi$_2$Se$_3$/1L PtSe$_2$ & 15  & 0.006 & 5.00  (CB) \\ 
                $-$          & 275 & 0.033 & 16.66 (VB) \\
2QL Bi$_2$Se$_3$/2L PtSe$_2$ & 11  & 0.005 & 4.40  (CB) \\
                $-$          & 278 & 0.033 & 16.84 (VB) \\ [1ex] 
\hline 
\end{tabular}
\label{table:table2} 
\end{table}

\begin{figure}[!htb]
\includegraphics[width=0.9\textwidth]{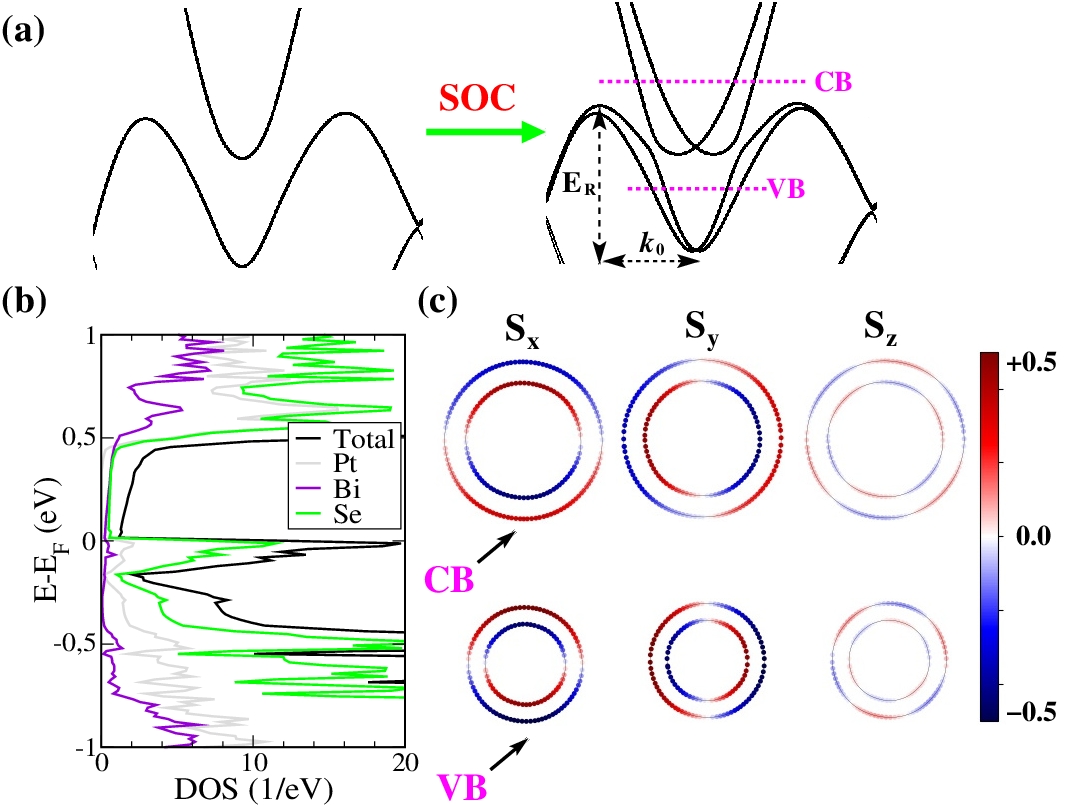}
\caption{For 1QL Bi$_2$Se$_3$/2L PtSe$_2$: (a) Stylized CB and VB regions around the Fermi level depicting band structure transformation under SOC. (b) Total and atom projected DOS showing orbital hybridization between constituent systems. (c) Fixed-energy (corresponding to magenta dotted line : 0.10 eV for the CB and -0.14 eV for the VB) contour plots of spin-components $s_x$,$s_y$ and $s_z$.}
\label{fig:fig6}
\end{figure}

To better describe these Rashba spin-splittings and associated spin-textures, Figure \ref{fig:fig6}(a) shows the bands changeover under the influence of SOC alongside the definition of $E_R$ and $k_0$ for the VB. Also, the total and atom projected DOS in Figure \ref{fig:fig6}(b) display significant changes in the orbital hybridization within the Bi$_2$Se$_3$ and PtSe$_2$ layers around the Fermi level as compared to Figure \ref{fig:fig2}, which is responsible for large SOC induced spin-splittings. We clarify this by plotting atomic-projected band structure of 1L Bi$_2$Se$_3$/2L PtSe$_2$ vdW heterostructures, shown in supplementary Figure \ref{fig:figs2prime}(c,d), where Se-atomic orbitals in the vicinity of the Fermi level hybridize and give rise to spin-splittings. In order to confirm the Rashba-shift along the momentum axis, we also give fixed-energy contour plots of spin-components $s_x$, $s_y$, and $s_z$ which corroborates our findings and reveal in-plane spin-components with minuscule out-of-plane contributions for both the CB and the VB as shown in Figure \ref{fig:fig6}(c). Comparing spin-splittings and Rashba-energies in the CB and the VB, we observe energy anisotropy in the spin texture similar to the case of graphene/TI heterostructure \cite{khokhriakov2018tailoring}. Since the optB86b-vdW functional is known to underestimate the band gap compared to experiment (and the use of a more accurate hybrid functional, such as HSE06, is prohibited due to the size of the models), it is anticipated that the transformation of type-II to type-III band alignment will occur at larger PtSe$_2$ thickness than 2L, which makes it possible to access the novel spintronics features predicted for these vdW heterostructures with few-layer Bi$_2$Se$_3$. Furthermore, low-temperature synthesis techniques are available for both systems, which also favors the formation of stacked vdW heterostructures.  

\section{Conclusion}

In summary, we employed density functional theory calculations to discuss layer-dependent electronic properties of few-layer Bi$_2$Se$_3$ and PtSe$_2$ van der Waals heterostructures. By varying the constituent layers, four different vdW heterostructures were constructed for which we provided details of structural, electronic and spintronics properties. It turned out that it is possible to simultaneously achieve type-II band alignment with layer-tunable band gaps and a complete 2D Rashba electron-gas spanning over a large energy interval of 0.9 eV for few-layer Bi$_2$Se$_3$/1L PtSe$_2$ vdW heterostructures. Fixed-energy contour plots of spin-components are presented to distinguish between Rashba- and hedgehog-like spin-textures of the CB and the VB originating from different constituent layers. By increasing the PtSe$_2$ thickness from 1L to 2L, we showed that type-II band alignment can be transformed to type-III. Moreover, we also revealed the coexistence of Rashba-like spin-splittings in both the CB (originating from Bi$_2$Se$_3$) and the VB (coming from PtSe$_2$) of few-layer Bi$_2$Se$_3$/2L PtSe$_2$ vdW heterostructures for which the corresponding spin-textures also support our findings. The role of inversion symmetry breaking, spin-orbit coupling and changes to the atomic orbital hybridization at the interface between few-layer Bi$_2$Se$_3$ and PtSe$_2$ is highlighted to understand and interpret the electronic dispersion. The electron localization function was used to analyze the interlayer binding and shows the contraction of the electron density within the Bi$_2$Se$_3$ and PtSe$_2$ layers. Our findings provide a unique avenue to manipulate the charge and spin degrees of freedom by exploiting few-layer Bi$_2$Se$_3$/PtSe$_2$ vdW heterostructures for potential electronic and spintronics applications.

\begin{acknowledgement}
Fruitful discussions with Muhammad Tahir and Aur\'{e}lien Manchon are greatly acknowledged. We thank Knut och Alice Wallenberg foundation, Kempestiftelserna and Interreg Nord for financial support. We also thank High Performance Computing Center North (HPC2N), National Supercomputer Center in Linköping (NSC), and the PDC Center for High Performance Computing for allocation of time and resources, through the Swedish National Infrastructure for Computing (SNIC).
\end{acknowledgement}

\begin{competing interests}
The Authors declare no competing financial or non-financial interests.
\end{competing interests}

\begin{data availability}
The data that support the findings of this study are available from the corresponding
author upon reasonable request.
\end{data availability}

\begin{author contribution}
S. Sattar performed the calculations, J. A. Larsson analysed the results.  All authors reviewed the manuscript. 
\end{author contribution}

\bibliography{main.bib}

\providecommand{\latin}[1]{#1}
\makeatletter
\providecommand{\doi}
  {\begingroup\let\do\@makeother\dospecials
  \catcode`\{=1 \catcode`\}=2 \doi@aux}
\providecommand{\doi@aux}[1]{\endgroup\texttt{#1}}
\makeatother
\providecommand*\mcitethebibliography{\thebibliography}
\csname @ifundefined\endcsname{endmcitethebibliography}
  {\let\endmcitethebibliography\endthebibliography}{}
\begin{mcitethebibliography}{52}
\providecommand*\natexlab[1]{#1}
\providecommand*\mciteSetBstSublistMode[1]{}
\providecommand*\mciteSetBstMaxWidthForm[2]{}
\providecommand*\mciteBstWouldAddEndPuncttrue
  {\def\EndOfBibitem{\unskip.}}
\providecommand*\mciteBstWouldAddEndPunctfalse
  {\let\EndOfBibitem\relax}
\providecommand*\mciteSetBstMidEndSepPunct[3]{}
\providecommand*\mciteSetBstSublistLabelBeginEnd[3]{}
\providecommand*\EndOfBibitem{}
\mciteSetBstSublistMode{f}
\mciteSetBstMaxWidthForm{subitem}{(\alph{mcitesubitemcount})}
\mciteSetBstSublistLabelBeginEnd
  {\mcitemaxwidthsubitemform\space}
  {\relax}
  {\relax}

\bibitem[Novoselov \latin{et~al.}(2004)Novoselov, Geim, Morozov, Jiang, Zhang,
  Dubonos, Grigorieva, and Firsov]{c1}
Novoselov,~K.~S.; Geim,~A.~K.; Morozov,~S.~V.; Jiang,~D.; Zhang,~Y.;
  Dubonos,~S.~V.; Grigorieva,~I.~V.; Firsov,~A.~A. Electric Field Effect in
  Atomically Thin Carbon Films. \emph{Science} \textbf{2004}, \emph{306},
  666--669\relax
\mciteBstWouldAddEndPuncttrue
\mciteSetBstMidEndSepPunct{\mcitedefaultmidpunct}
{\mcitedefaultendpunct}{\mcitedefaultseppunct}\relax
\EndOfBibitem
\bibitem[Novoselov \latin{et~al.}(2005)Novoselov, Geim, Morozov, Jiang,
  Katsnelson, Grigorieva, Dubonos, and Firsov]{c2}
Novoselov,~K.~S.; Geim,~A.~K.; Morozov,~S.; Jiang,~D.; Katsnelson,~M.~I.;
  Grigorieva,~I.; Dubonos,~S.; Firsov,~A.~A. Two-Dimensional Gas of Massless
  Dirac Fermions in Graphene. \emph{Nature} \textbf{2005}, \emph{438},
  197--200\relax
\mciteBstWouldAddEndPuncttrue
\mciteSetBstMidEndSepPunct{\mcitedefaultmidpunct}
{\mcitedefaultendpunct}{\mcitedefaultseppunct}\relax
\EndOfBibitem
\bibitem[Zhang \latin{et~al.}(2005)Zhang, Tan, Stormer, and Kim]{c3}
Zhang,~Y.; Tan,~Y.-W.; Stormer,~H.~L.; Kim,~P. Experimental Observation of the
  Quantum Hall Effect and Berry's Phase in Graphene. \emph{Nature}
  \textbf{2005}, \emph{438}, 201--204\relax
\mciteBstWouldAddEndPuncttrue
\mciteSetBstMidEndSepPunct{\mcitedefaultmidpunct}
{\mcitedefaultendpunct}{\mcitedefaultseppunct}\relax
\EndOfBibitem
\bibitem[Novoselov \latin{et~al.}(2005)Novoselov, Jiang, Schedin, Booth,
  Khotkevich, Morozov, and Geim]{mos21}
Novoselov,~K.~S.; Jiang,~D.; Schedin,~F.; Booth,~T.~J.; Khotkevich,~V.~V.;
  Morozov,~S.~V.; Geim,~A.~K. Two-Dimensional Atomic Crystals.
  \emph{Proceedings of the National Academy of Sciences} \textbf{2005},
  \emph{102}, 10451--10453\relax
\mciteBstWouldAddEndPuncttrue
\mciteSetBstMidEndSepPunct{\mcitedefaultmidpunct}
{\mcitedefaultendpunct}{\mcitedefaultseppunct}\relax
\EndOfBibitem
\bibitem[Mak \latin{et~al.}(2010)Mak, Lee, Hone, Shan, and Heinz]{mos22}
Mak,~K.~F.; Lee,~C.; Hone,~J.; Shan,~J.; Heinz,~T.~F. Atomically Thin
  ${\mathrm{MoS}}_{2}$: A New Direct-Gap Semiconductor. \emph{Phys. Rev. Lett.}
  \textbf{2010}, \emph{105}, 136805\relax
\mciteBstWouldAddEndPuncttrue
\mciteSetBstMidEndSepPunct{\mcitedefaultmidpunct}
{\mcitedefaultendpunct}{\mcitedefaultseppunct}\relax
\EndOfBibitem
\bibitem[Radisavljevic \latin{et~al.}(2011)Radisavljevic, Radenovic, Brivio,
  Giacometti, and Kis]{mos23}
Radisavljevic,~B.; Radenovic,~A.; Brivio,~J.; Giacometti,~V.; Kis,~A.
  Single-layer MoS${}_{2}$ Transistors. \emph{Nature Nanotechnology}
  \textbf{2011}, \emph{6}, 147\relax
\mciteBstWouldAddEndPuncttrue
\mciteSetBstMidEndSepPunct{\mcitedefaultmidpunct}
{\mcitedefaultendpunct}{\mcitedefaultseppunct}\relax
\EndOfBibitem
\bibitem[Mas-Ballesté \latin{et~al.}(2011)Mas-Ballesté, Gómez-Navarro,
  Gómez-Herrero, and Zamora]{vdw1}
Mas-Ballesté,~R.; Gómez-Navarro,~C.; Gómez-Herrero,~J.; Zamora,~F. 2D
  Materials: To Graphene and Beyond. \emph{Nanoscale} \textbf{2011}, \emph{3},
  20--30\relax
\mciteBstWouldAddEndPuncttrue
\mciteSetBstMidEndSepPunct{\mcitedefaultmidpunct}
{\mcitedefaultendpunct}{\mcitedefaultseppunct}\relax
\EndOfBibitem
\bibitem[Geim and Grigorieva(2013)Geim, and Grigorieva]{vdw2}
Geim,~A.~K.; Grigorieva,~I.~V. Van der Waals Heterostructures. \emph{Nature}
  \textbf{2013}, \emph{499}, 419--425\relax
\mciteBstWouldAddEndPuncttrue
\mciteSetBstMidEndSepPunct{\mcitedefaultmidpunct}
{\mcitedefaultendpunct}{\mcitedefaultseppunct}\relax
\EndOfBibitem
\bibitem[Xiang \latin{et~al.}(2020)Xiang, Inoue, Zheng, Kumamoto, Qian, Sato,
  Liu, Tang, Gokhale, Guo, Hisama, Yotsumoto, Ogamoto, Arai, Kobayashi, Zhang,
  Hou, Anisimov, Maruyama, Miyata, Okada, Chiashi, Li, Kong, Kauppinen,
  Ikuhara, Suenaga, and Maruyama]{vdw3}
Xiang,~R.; Inoue,~T.; Zheng,~Y.; Kumamoto,~A.; Qian,~Y.; Sato,~Y.; Liu,~M.;
  Tang,~D.; Gokhale,~D.; Guo,~J.; Hisama,~K.; Yotsumoto,~S.; Ogamoto,~T.;
  Arai,~H.; Kobayashi,~Y.; Zhang,~H.; Hou,~B.; Anisimov,~A.; Maruyama,~M.;
  Miyata,~Y.; Okada,~S.; Chiashi,~S.; Li,~Y.; Kong,~J.; Kauppinen,~E.~I.;
  Ikuhara,~Y.; Suenaga,~K.; Maruyama,~S. One-Dimensional van der Waals
  Heterostructures. \emph{Science} \textbf{2020}, \emph{367}, 537--542\relax
\mciteBstWouldAddEndPuncttrue
\mciteSetBstMidEndSepPunct{\mcitedefaultmidpunct}
{\mcitedefaultendpunct}{\mcitedefaultseppunct}\relax
\EndOfBibitem
\bibitem[Ciarrocchi \latin{et~al.}(2018)Ciarrocchi, Avsar, Ovchinnikov, and
  Kis]{bandgap1}
Ciarrocchi,~A.; Avsar,~A.; Ovchinnikov,~D.; Kis,~A. Thickness-Modulated
  Metal-to-Semiconductor Transformation in a Transition Metal Dichalcogenide.
  \emph{Nature Communications} \textbf{2018}, \emph{9}, 1--6\relax
\mciteBstWouldAddEndPuncttrue
\mciteSetBstMidEndSepPunct{\mcitedefaultmidpunct}
{\mcitedefaultendpunct}{\mcitedefaultseppunct}\relax
\EndOfBibitem
\bibitem[Zhao \latin{et~al.}(2017)Zhao, Qiao, Yu, Yu, Xu, Lau, Zhou, Liu, Wang,
  Ji, and Chai]{mobi1}
Zhao,~Y.; Qiao,~J.; Yu,~Z.; Yu,~P.; Xu,~K.; Lau,~S.~P.; Zhou,~W.; Liu,~Z.;
  Wang,~X.; Ji,~W.; Chai,~Y. High-Electron-Mobility and Air-Stable 2D Layered
  PtSe${}_{2}$ FETs. \emph{Advanced Materials} \textbf{2017}, \emph{29},
  1604230\relax
\mciteBstWouldAddEndPuncttrue
\mciteSetBstMidEndSepPunct{\mcitedefaultmidpunct}
{\mcitedefaultendpunct}{\mcitedefaultseppunct}\relax
\EndOfBibitem
\bibitem[Okogbue \latin{et~al.}(2019)Okogbue, Han, Ko, Chung, Ma, Shawkat, Kim,
  Kim, Ji, Oh, Zhai, Lee, and Jung]{str1}
Okogbue,~E.; Han,~S.~S.; Ko,~T.-J.; Chung,~H.-S.; Ma,~J.; Shawkat,~M.~S.;
  Kim,~J.~H.; Kim,~J.~H.; Ji,~E.; Oh,~K.~H.; Zhai,~L.; Lee,~G.-H.; Jung,~Y.
  Multifunctional Two-Dimensional PtSe${}_{2}$-Layer Kirigami Conductors with
  2000\% Stretchability and Metallic-to-Semiconducting Tunability. \emph{Nano
  Letters} \textbf{2019}, \emph{19}, 7598--7607\relax
\mciteBstWouldAddEndPuncttrue
\mciteSetBstMidEndSepPunct{\mcitedefaultmidpunct}
{\mcitedefaultendpunct}{\mcitedefaultseppunct}\relax
\EndOfBibitem
\bibitem[Sun \latin{et~al.}(2020)Sun, Zhang, Li, Zheng, Wu, Li, Ding, Lv, and
  Tao]{cat1}
Sun,~X.; Zhang,~H.; Li,~X.; Zheng,~Y.-Z.; Wu,~J.; Li,~N.; Ding,~H.; Lv,~X.;
  Tao,~X. An Efficient and Extremely Stable Photocatalytic PtSe${}_{2}$/FTO
  Thin Film for Water Splitting. \emph{Energy Technology} \textbf{2020},
  \emph{8}, 1900903\relax
\mciteBstWouldAddEndPuncttrue
\mciteSetBstMidEndSepPunct{\mcitedefaultmidpunct}
{\mcitedefaultendpunct}{\mcitedefaultseppunct}\relax
\EndOfBibitem
\bibitem[Yim \latin{et~al.}(2016)Yim, Lee, McEvoy, O’Brien, Riazimehr,
  Berner, Cullen, Kotakoski, Meyer, Lemme, and Duesberg]{opto1}
Yim,~C.; Lee,~K.; McEvoy,~N.; O’Brien,~M.; Riazimehr,~S.; Berner,~N.~C.;
  Cullen,~C.~P.; Kotakoski,~J.; Meyer,~J.~C.; Lemme,~M.~C.; Duesberg,~G.~S.
  High-Performance Hybrid Electronic Devices from Layered PtSe${}_{2}$ Films
  Grown at Low Temperature. \emph{ACS Nano} \textbf{2016}, \emph{10},
  9550--9558\relax
\mciteBstWouldAddEndPuncttrue
\mciteSetBstMidEndSepPunct{\mcitedefaultmidpunct}
{\mcitedefaultendpunct}{\mcitedefaultseppunct}\relax
\EndOfBibitem
\bibitem[Wang \latin{et~al.}(2015)Wang, Li, Yao, Song, Sun, Pan, Ren, Li,
  Okunishi, Wang, Wang, Shao, Zhang, Yang, Schwier, Iwasawa, Shimada,
  Taniguchi, Cheng, Zhou, Du, Pennycook, Pantelides, and Gao]{pt111}
Wang,~Y.; Li,~L.; Yao,~W.; Song,~S.; Sun,~J.~T.; Pan,~J.; Ren,~X.; Li,~C.;
  Okunishi,~E.; Wang,~Y.-Q.; Wang,~E.; Shao,~Y.; Zhang,~Y.~Y.; Yang,~H.-T.;
  Schwier,~E.~F.; Iwasawa,~H.; Shimada,~K.; Taniguchi,~M.; Cheng,~Z.; Zhou,~S.;
  Du,~S.; Pennycook,~S.~J.; Pantelides,~S.~T.; Gao,~H.-J. Monolayer
  PtSe${}_{2}$, a New Semiconducting Transition-Metal-Dichalcogenide,
  Epitaxially Grown by Direct Selenization of Pt. \emph{Nano Letters}
  \textbf{2015}, \emph{15}, 4013--4018\relax
\mciteBstWouldAddEndPuncttrue
\mciteSetBstMidEndSepPunct{\mcitedefaultmidpunct}
{\mcitedefaultendpunct}{\mcitedefaultseppunct}\relax
\EndOfBibitem
\bibitem[Ansari \latin{et~al.}(2019)Ansari, Monaghan, McEvoy, Coile{\'a}in,
  Cullen, Lin, Siris, Stimpel-Lindner, Burke, Mirabelli, Duffy, Caruso, Nagle,
  Duesberg, Hurley, and Gity]{si}
Ansari,~L.; Monaghan,~S.; McEvoy,~N.; Coile{\'a}in,~C.~{\'O}.; Cullen,~C.~P.;
  Lin,~J.; Siris,~R.; Stimpel-Lindner,~T.; Burke,~K.~F.; Mirabelli,~G.;
  Duffy,~R.; Caruso,~E.; Nagle,~R.~E.; Duesberg,~G.~S.; Hurley,~P.~K.; Gity,~F.
  Quantum Confinement-Induced Semimetal-to-Semiconductor Evolution in
  Large-Area Ultra-Thin PtSe${}_{2}$ Films Grown at 400 \degree C. \emph{npj 2D
  Materials and Applications} \textbf{2019}, \emph{3}, 1--8\relax
\mciteBstWouldAddEndPuncttrue
\mciteSetBstMidEndSepPunct{\mcitedefaultmidpunct}
{\mcitedefaultendpunct}{\mcitedefaultseppunct}\relax
\EndOfBibitem
\bibitem[Wang \latin{et~al.}(2016)Wang, Li, Besenbacher, and Dong]{saphire}
Wang,~Z.; Li,~Q.; Besenbacher,~F.; Dong,~M. Facile Synthesis of Single Crystal
  PtSe${}_{2}$ Nanosheets for Nanoscale Electronics. \emph{Advanced Materials}
  \textbf{2016}, \emph{28}, 10224--10229\relax
\mciteBstWouldAddEndPuncttrue
\mciteSetBstMidEndSepPunct{\mcitedefaultmidpunct}
{\mcitedefaultendpunct}{\mcitedefaultseppunct}\relax
\EndOfBibitem
\bibitem[Yan \latin{et~al.}(2017)Yan, Wang, Zhou, Zhang, Zhang, Zhang, Yao, Lu,
  Yang, Wu, Yoshikawa, Miyamoto, Okuda, Wu, Yu, Duan, and Zhou]{graphene}
Yan,~M.; Wang,~E.; Zhou,~X.; Zhang,~G.; Zhang,~H.; Zhang,~K.; Yao,~W.; Lu,~N.;
  Yang,~S.; Wu,~S.; Yoshikawa,~T.; Miyamoto,~K.; Okuda,~T.; Wu,~Y.; Yu,~P.;
  Duan,~W.; Zhou,~S. High Quality Atomically Thin PtSe${}_{2}$ Films Grown by
  Molecular Beam Epitaxy. \emph{2D Materials} \textbf{2017}, \emph{4},
  045015\relax
\mciteBstWouldAddEndPuncttrue
\mciteSetBstMidEndSepPunct{\mcitedefaultmidpunct}
{\mcitedefaultendpunct}{\mcitedefaultseppunct}\relax
\EndOfBibitem
\bibitem[{Xiong} \latin{et~al.}(2020){Xiong}, {Hilse}, {Li}, {Göritz},
  {Lisker}, {Wietstruck}, {Kaynak}, {Engel-Herbert}, {Madjar}, and
  {Hwang}]{mosfet}
{Xiong},~K.; {Hilse},~M.; {Li},~L.; {Göritz},~A.; {Lisker},~M.;
  {Wietstruck},~M.; {Kaynak},~M.; {Engel-Herbert},~R.; {Madjar},~A.;
  {Hwang},~J. C.~M. Large-Scale Fabrication of Submicrometer-Gate-Length
  MOSFETs With a Trilayer PtSe${}_{2}$ Channel Grown by Molecular Beam Epitaxy.
  \emph{IEEE Transactions on Electron Devices} \textbf{2020}, \emph{67},
  796--801\relax
\mciteBstWouldAddEndPuncttrue
\mciteSetBstMidEndSepPunct{\mcitedefaultmidpunct}
{\mcitedefaultendpunct}{\mcitedefaultseppunct}\relax
\EndOfBibitem
\bibitem[Yao \latin{et~al.}(2017)Yao, Wang, Huang, Deng, Yan, Zhang, Miyamoto,
  Okuda, Li, Wang, Gao, Liu, Duan, and Zhou]{spinptse2}
Yao,~W.; Wang,~E.; Huang,~H.; Deng,~K.; Yan,~M.; Zhang,~K.; Miyamoto,~K.;
  Okuda,~T.; Li,~L.; Wang,~Y.; Gao,~H.; Liu,~C.; Duan,~W.; Zhou,~S. Direct
  Observation of Spin-Layer Locking by Local Rashba Effect in Monolayer
  Semiconducting PtSe${}_{2}$ Film. \emph{Nature Communications} \textbf{2017},
  \emph{8}, 1--6\relax
\mciteBstWouldAddEndPuncttrue
\mciteSetBstMidEndSepPunct{\mcitedefaultmidpunct}
{\mcitedefaultendpunct}{\mcitedefaultseppunct}\relax
\EndOfBibitem
\bibitem[Zhou \latin{et~al.}(2019)Zhou, Kong, Sekhar, Lin, Le~Goualher, Xu,
  Wang, Chen, Zhou, Zhu, Lu, Liu, Tang, Guo, Zhu, Cheng, Yu, Suenaga, Sun, Ji,
  and Liu]{ptvdw}
Zhou,~J.; Kong,~X.; Sekhar,~M.~C.; Lin,~J.; Le~Goualher,~F.; Xu,~R.; Wang,~X.;
  Chen,~Y.; Zhou,~Y.; Zhu,~C.; Lu,~W.; Liu,~F.; Tang,~B.; Guo,~Z.; Zhu,~C.;
  Cheng,~Z.; Yu,~T.; Suenaga,~K.; Sun,~D.; Ji,~W.; Liu,~Z. Epitaxial Synthesis
  of Monolayer PtSe${}_{2}$ Single Crystal on MoSe${}_{2}$ with Strong
  Interlayer Coupling. \emph{ACS Nano} \textbf{2019}, \emph{13},
  10929--10938\relax
\mciteBstWouldAddEndPuncttrue
\mciteSetBstMidEndSepPunct{\mcitedefaultmidpunct}
{\mcitedefaultendpunct}{\mcitedefaultseppunct}\relax
\EndOfBibitem
\bibitem[Li \latin{et~al.}(2014)Li, Van‘t~Erve, Robinson, Liu, Li, and
  Jonker]{li2014electrical}
Li,~C.; Van‘t~Erve,~O.; Robinson,~J.; Liu,~Y.; Li,~L.; Jonker,~B. Electrical
  detection of charge-current-induced spin polarization due to spin-momentum
  locking in Bi${}_{2}$Se${}_{3}$. \emph{Nature nanotechnology} \textbf{2014},
  \emph{9}, 218\relax
\mciteBstWouldAddEndPuncttrue
\mciteSetBstMidEndSepPunct{\mcitedefaultmidpunct}
{\mcitedefaultendpunct}{\mcitedefaultseppunct}\relax
\EndOfBibitem
\bibitem[Dankert \latin{et~al.}(2015)Dankert, Geurs, Kamalakar, Charpentier,
  and Dash]{dankert2015room}
Dankert,~A.; Geurs,~J.; Kamalakar,~M.~V.; Charpentier,~S.; Dash,~S.~P. Room
  temperature electrical detection of spin polarized currents in topological
  insulators. \emph{Nano letters} \textbf{2015}, \emph{15}, 7976--7981\relax
\mciteBstWouldAddEndPuncttrue
\mciteSetBstMidEndSepPunct{\mcitedefaultmidpunct}
{\mcitedefaultendpunct}{\mcitedefaultseppunct}\relax
\EndOfBibitem
\bibitem[Dankert \latin{et~al.}(2018)Dankert, Bhaskar, Khokhriakov, Rodrigues,
  Karpiak, Kamalakar, Charpentier, Garate, and Dash]{dankert2018origin}
Dankert,~A.; Bhaskar,~P.; Khokhriakov,~D.; Rodrigues,~I.~H.; Karpiak,~B.;
  Kamalakar,~M.~V.; Charpentier,~S.; Garate,~I.; Dash,~S.~P. Origin and
  evolution of surface spin current in topological insulators. \emph{Physical
  Review B} \textbf{2018}, \emph{97}, 125414\relax
\mciteBstWouldAddEndPuncttrue
\mciteSetBstMidEndSepPunct{\mcitedefaultmidpunct}
{\mcitedefaultendpunct}{\mcitedefaultseppunct}\relax
\EndOfBibitem
\bibitem[jam()]{jamali2015giant}
Giant spin pumping and inverse spin Hall effect in the presence of surface and
  bulk spin- orbit coupling of topological insulator Bi${}_{2}$Se${}_{3}$.
  \relax
\mciteBstWouldAddEndPunctfalse
\mciteSetBstMidEndSepPunct{\mcitedefaultmidpunct}
{}{\mcitedefaultseppunct}\relax
\EndOfBibitem
\bibitem[Zhang \latin{et~al.}(2010)Zhang, He, Chang, Song, Wang, Chen, Jia,
  Fang, Dai, Shan, \latin{et~al.} others]{bise1}
Zhang,~Y.; He,~K.; Chang,~C.-Z.; Song,~C.-L.; Wang,~L.-L.; Chen,~X.;
  Jia,~J.-F.; Fang,~Z.; Dai,~X.; Shan,~W.-Y., \latin{et~al.}  Crossover of the
  Three-Dimensional Topological Insulator Bi${}_{2}$Se${}_{3}$ to the
  Two-Dimensional Limit. \emph{Nature Physics} \textbf{2010}, \emph{6},
  584--588\relax
\mciteBstWouldAddEndPuncttrue
\mciteSetBstMidEndSepPunct{\mcitedefaultmidpunct}
{\mcitedefaultendpunct}{\mcitedefaultseppunct}\relax
\EndOfBibitem
\bibitem[Zhang \latin{et~al.}(2009)Zhang, Liu, Qi, Dai, Fang, and Zhang]{bise2}
Zhang,~H.; Liu,~C.-X.; Qi,~X.-L.; Dai,~X.; Fang,~Z.; Zhang,~S.-C. Topological
  Insulators in Bi${}_{2}$Se${}_{3}$, Bi${}_{2}$Te${}_{3}$ and
  Sb${}_{2}$Te${}_{3}$ With a Single Dirac Cone on the Surface. \emph{Nature
  Physics} \textbf{2009}, \emph{5}, 438--442\relax
\mciteBstWouldAddEndPuncttrue
\mciteSetBstMidEndSepPunct{\mcitedefaultmidpunct}
{\mcitedefaultendpunct}{\mcitedefaultseppunct}\relax
\EndOfBibitem
\bibitem[Hsieh \latin{et~al.}(2009)Hsieh, Xia, Qian, Wray, Dil, Meier,
  Osterwalder, Patthey, Checkelsky, Ong, Fedorov, Lin, Bansil, Grauer, Hor,
  Cava, and Hasan]{bise3}
Hsieh,~D.; Xia,~Y.; Qian,~D.; Wray,~L.; Dil,~J.~H.; Meier,~F.; Osterwalder,~J.;
  Patthey,~L.; Checkelsky,~J.~G.; Ong,~N.~P.; Fedorov,~A.~V.; Lin,~H.;
  Bansil,~A.; Grauer,~D.; Hor,~Y.~S.; Cava,~R.~J.; Hasan,~M.~Z. A Tunable
  Topological Insulator in the Spin Helical Dirac Transport Regime.
  \emph{Nature} \textbf{2009}, \emph{460}, 1101--1105\relax
\mciteBstWouldAddEndPuncttrue
\mciteSetBstMidEndSepPunct{\mcitedefaultmidpunct}
{\mcitedefaultendpunct}{\mcitedefaultseppunct}\relax
\EndOfBibitem
\bibitem[Wang \latin{et~al.}(2012)Wang, Liu, Xu, Yang, Miao, Yao, Gao, Shen,
  Ma, Chen, Xu, Liu, Zhang, Qian, Jia, and Xue]{sup-top}
Wang,~M.-X.; Liu,~C.; Xu,~J.-P.; Yang,~F.; Miao,~L.; Yao,~M.-Y.; Gao,~C.~L.;
  Shen,~C.; Ma,~X.; Chen,~X.; Xu,~Z.-A.; Liu,~Y.; Zhang,~S.-C.; Qian,~D.;
  Jia,~J.-F.; Xue,~Q.-K. The Coexistence of Superconductivity and Topological
  Order in the Bi${}_{2}$Se${}_{3}$ Thin Films. \emph{Science} \textbf{2012},
  \emph{336}, 52--55\relax
\mciteBstWouldAddEndPuncttrue
\mciteSetBstMidEndSepPunct{\mcitedefaultmidpunct}
{\mcitedefaultendpunct}{\mcitedefaultseppunct}\relax
\EndOfBibitem
\bibitem[Xu \latin{et~al.}(2012)Xu, Neupane, Liu, Zhang, Richardella, Wray,
  Alidoust, Leandersson, Balasubramanian, S{\'a}nchez-Barriga, Rader, Landolt,
  Slomski, Dil, Osterwalder, Chang, Jeng, Bansil, Samarth, and Hasan]{berry}
Xu,~S.-Y.; Neupane,~M.; Liu,~C.; Zhang,~D.; Richardella,~A.; Wray,~L.~A.;
  Alidoust,~N.; Leandersson,~M.; Balasubramanian,~T.; S{\'a}nchez-Barriga,~J.;
  Rader,~O.; Landolt,~G.; Slomski,~B.; Dil,~J.~H.; Osterwalder,~J.;
  Chang,~T.-R.; Jeng,~H.-T.; Bansil,~A.; Samarth,~N.; Hasan,~M.~Z. Hedgehog
  Spin Texture and Berry’s Phase Tuning in a Magnetic Topological Insulator.
  \emph{Nature Physics} \textbf{2012}, \emph{8}, 616--622\relax
\mciteBstWouldAddEndPuncttrue
\mciteSetBstMidEndSepPunct{\mcitedefaultmidpunct}
{\mcitedefaultendpunct}{\mcitedefaultseppunct}\relax
\EndOfBibitem
\bibitem[Guo \latin{et~al.}(2016)Guo, Wang, Xu, Huang, Zang, Liu, Duan, Gan,
  Zhang, He, Ma, Xue, and Wang]{thermo}
Guo,~M.; Wang,~Z.; Xu,~Y.; Huang,~H.; Zang,~Y.; Liu,~C.; Duan,~W.; Gan,~Z.;
  Zhang,~S.-C.; He,~K.; Ma,~X.; Xue,~Q.; Wang,~Y. Tuning Thermoelectricity in a
  Bi${}_{2}$Se${}_{3}$ Topological Insulator via Varied Film Thickness.
  \emph{New Journal of Physics} \textbf{2016}, \emph{18}, 015008\relax
\mciteBstWouldAddEndPuncttrue
\mciteSetBstMidEndSepPunct{\mcitedefaultmidpunct}
{\mcitedefaultendpunct}{\mcitedefaultseppunct}\relax
\EndOfBibitem
\bibitem[Glinka \latin{et~al.}(2013)Glinka, Babakiray, Johnson, Bristow,
  Holcomb, and Lederman]{ultfast}
Glinka,~Y.~D.; Babakiray,~S.; Johnson,~T.~A.; Bristow,~A.~D.; Holcomb,~M.~B.;
  Lederman,~D. Ultrafast Carrier Dynamics in Thin-Films of the Topological
  Insulator Bi${}_{2}$Se${}_{3}$. \emph{Applied Physics Letters} \textbf{2013},
  \emph{103}, 151903\relax
\mciteBstWouldAddEndPuncttrue
\mciteSetBstMidEndSepPunct{\mcitedefaultmidpunct}
{\mcitedefaultendpunct}{\mcitedefaultseppunct}\relax
\EndOfBibitem
\bibitem[Benia \latin{et~al.}(2013)Benia, Yaresko, Schnyder, Henk, Lin, Kern,
  and Ast]{bise-r}
Benia,~H.~M.; Yaresko,~A.; Schnyder,~A.~P.; Henk,~J.; Lin,~C.~T.; Kern,~K.;
  Ast,~C.~R. Origin of Rashba Splitting in the Quantized Subbands at the
  Bi${}_{2}$Se${}_{3}$ surface. \emph{Phys. Rev. B} \textbf{2013}, \emph{88},
  081103\relax
\mciteBstWouldAddEndPuncttrue
\mciteSetBstMidEndSepPunct{\mcitedefaultmidpunct}
{\mcitedefaultendpunct}{\mcitedefaultseppunct}\relax
\EndOfBibitem
\bibitem[Zhang and Schwingenschl\"ogl(2018)Zhang, and
  Schwingenschl\"ogl]{alpha1}
Zhang,~Q.; Schwingenschl\"ogl,~U. Rashba Effect and Enriched Spin-Valley
  Coupling in $\mathrm{Ga}X$/$M{X}_{2}$ ($M$ = Mo, W; $X$ = S, Se, Te)
  Heterostructures. \emph{Phys. Rev. B} \textbf{2018}, \emph{97}, 155415\relax
\mciteBstWouldAddEndPuncttrue
\mciteSetBstMidEndSepPunct{\mcitedefaultmidpunct}
{\mcitedefaultendpunct}{\mcitedefaultseppunct}\relax
\EndOfBibitem
\bibitem[Wang and Jeng(2017)Wang, and Jeng]{bise-r2}
Wang,~T.-H.; Jeng,~H.-T. Wide-Range Ideal 2D Rashba Electron Gas with Large
  Spin Splitting in Bi${}_{2}$Se${}_{3}$/MoTe${}_{2}$ Heterostructure.
  \emph{npj Computational Materials} \textbf{2017}, \emph{3}, 1--6\relax
\mciteBstWouldAddEndPuncttrue
\mciteSetBstMidEndSepPunct{\mcitedefaultmidpunct}
{\mcitedefaultendpunct}{\mcitedefaultseppunct}\relax
\EndOfBibitem
\bibitem[Bl\"ochl(1994)]{paw1}
Bl\"ochl,~P.~E. Projector Augmented-Wave Method. \emph{Phys. Rev. B}
  \textbf{1994}, \emph{50}, 17953--17979\relax
\mciteBstWouldAddEndPuncttrue
\mciteSetBstMidEndSepPunct{\mcitedefaultmidpunct}
{\mcitedefaultendpunct}{\mcitedefaultseppunct}\relax
\EndOfBibitem
\bibitem[Kresse and Furthm\"uller(1996)Kresse, and Furthm\"uller]{paw2}
Kresse,~G.; Furthm\"uller,~J. Efficient Iterative Schemes for Ab Initio
  Total-Energy Calculations Using a Plane-Wave Basis Set. \emph{Phys. Rev. B}
  \textbf{1996}, \emph{54}, 11169--11186\relax
\mciteBstWouldAddEndPuncttrue
\mciteSetBstMidEndSepPunct{\mcitedefaultmidpunct}
{\mcitedefaultendpunct}{\mcitedefaultseppunct}\relax
\EndOfBibitem
\bibitem[Kresse and Joubert(1999)Kresse, and Joubert]{vasp}
Kresse,~G.; Joubert,~D. From Ultrasoft Pseudopotentials to the Projector
  Augmented-Wave Method. \emph{Phys. Rev. B} \textbf{1999}, \emph{59},
  1758--1775\relax
\mciteBstWouldAddEndPuncttrue
\mciteSetBstMidEndSepPunct{\mcitedefaultmidpunct}
{\mcitedefaultendpunct}{\mcitedefaultseppunct}\relax
\EndOfBibitem
\bibitem[Klime{\v{s}} \latin{et~al.}(2009)Klime{\v{s}}, Bowler, and
  Michaelides]{optb86}
Klime{\v{s}},~J.; Bowler,~D.~R.; Michaelides,~A. Chemical Accuracy For the Van
  der Waals Density Functional. \emph{Journal of Physics: Condensed Matter}
  \textbf{2009}, \emph{22}, 022201\relax
\mciteBstWouldAddEndPuncttrue
\mciteSetBstMidEndSepPunct{\mcitedefaultmidpunct}
{\mcitedefaultendpunct}{\mcitedefaultseppunct}\relax
\EndOfBibitem
\bibitem[Klime{\v{s}} \latin{et~al.}(2011)Klime{\v{s}}, Bowler, and
  Michaelides]{optb86-2}
Klime{\v{s}},~J.; Bowler,~D.~R.; Michaelides,~A. Van der Waals Density
  Functionals Applied to Solids. \emph{Phys. Rev. B} \textbf{2011}, \emph{83},
  195131\relax
\mciteBstWouldAddEndPuncttrue
\mciteSetBstMidEndSepPunct{\mcitedefaultmidpunct}
{\mcitedefaultendpunct}{\mcitedefaultseppunct}\relax
\EndOfBibitem
\bibitem[Peng \latin{et~al.}(2016)Peng, Wang, Sa, Wu, and
  Sun]{peng2016electronic}
Peng,~Q.; Wang,~Z.; Sa,~B.; Wu,~B.; Sun,~Z. Electronic Structures and Enhanced
  Optical Properties of Blue Phosphorene/Transition Metal Dichalcogenides van
  der Waals Heterostructures. \emph{Scientific reports} \textbf{2016},
  \emph{6}, 31994\relax
\mciteBstWouldAddEndPuncttrue
\mciteSetBstMidEndSepPunct{\mcitedefaultmidpunct}
{\mcitedefaultendpunct}{\mcitedefaultseppunct}\relax
\EndOfBibitem
\bibitem[Sharma and Schwingenschlögl(2016)Sharma, and
  Schwingenschlögl]{bi2te3}
Sharma,~S.; Schwingenschlögl,~U. Thermoelectric Response in Single Quintuple
  Layer Bi${}_{2}$Te${}_{3}$. \emph{ACS Energy Letters} \textbf{2016},
  \emph{1}, 875--879\relax
\mciteBstWouldAddEndPuncttrue
\mciteSetBstMidEndSepPunct{\mcitedefaultmidpunct}
{\mcitedefaultendpunct}{\mcitedefaultseppunct}\relax
\EndOfBibitem
\bibitem[Herath \latin{et~al.}(2020)Herath, Tavadze, He, Bousquet, Singh,
  Muñoz, and Romero]{pyprocar}
Herath,~U.; Tavadze,~P.; He,~X.; Bousquet,~E.; Singh,~S.; Muñoz,~F.;
  Romero,~A.~H. PyProcar: A Python Library for Electronic Structure
  Pre/Post-Processing. \emph{Computer Physics Communications} \textbf{2020},
  \emph{251}, 107080\relax
\mciteBstWouldAddEndPuncttrue
\mciteSetBstMidEndSepPunct{\mcitedefaultmidpunct}
{\mcitedefaultendpunct}{\mcitedefaultseppunct}\relax
\EndOfBibitem
\bibitem[{Hunter}(2007)]{matplotlib}
{Hunter},~J.~D. Matplotlib: A 2D Graphics Environment. \emph{Computing in
  Science Engineering} \textbf{2007}, \emph{9}, 90--95\relax
\mciteBstWouldAddEndPuncttrue
\mciteSetBstMidEndSepPunct{\mcitedefaultmidpunct}
{\mcitedefaultendpunct}{\mcitedefaultseppunct}\relax
\EndOfBibitem
\bibitem[Koumpouras and Larsson(2020)Koumpouras, and Larsson]{and1}
Koumpouras,~K.; Larsson,~J.~A. Distinguishing Between Chemical Bonding and
  Physical Binding Using Electron Localization Function (ELF). \emph{Journal of
  Physics: Condensed Matter} \textbf{2020}, \emph{32}, 315502\relax
\mciteBstWouldAddEndPuncttrue
\mciteSetBstMidEndSepPunct{\mcitedefaultmidpunct}
{\mcitedefaultendpunct}{\mcitedefaultseppunct}\relax
\EndOfBibitem
\bibitem[Sajjad \latin{et~al.}(2018)Sajjad, Singh, and
  Schwingenschlögl]{exciton}
Sajjad,~M.; Singh,~N.; Schwingenschlögl,~U. Strongly Bound Excitons in
  Monolayer PtS${}_{2}$ and PtSe${}_{2}$. \emph{Applied Physics Letters}
  \textbf{2018}, \emph{112}, 043101\relax
\mciteBstWouldAddEndPuncttrue
\mciteSetBstMidEndSepPunct{\mcitedefaultmidpunct}
{\mcitedefaultendpunct}{\mcitedefaultseppunct}\relax
\EndOfBibitem
\bibitem[Kung \latin{et~al.}(2019)Kung, Goyal, Maslov, Wang, Lee, Kemper,
  Cheong, and Blumberg]{chiral}
Kung,~H.-H.; Goyal,~A.~P.; Maslov,~D.~L.; Wang,~X.; Lee,~A.; Kemper,~A.~F.;
  Cheong,~S.-W.; Blumberg,~G. Observation of Chiral Surface Excitons in a
  Topological Insulator Bi${}_{2}$Se${}_{3}$. \emph{Proceedings of the National
  Academy of Sciences} \textbf{2019}, \emph{116}, 4006--4011\relax
\mciteBstWouldAddEndPuncttrue
\mciteSetBstMidEndSepPunct{\mcitedefaultmidpunct}
{\mcitedefaultendpunct}{\mcitedefaultseppunct}\relax
\EndOfBibitem
\bibitem[Khokhriakov \latin{et~al.}(2020)Khokhriakov, Hoque, Karpiak, and
  Dash]{khokhriakov2020gate}
Khokhriakov,~D.; Hoque,~A.~M.; Karpiak,~B.; Dash,~S.~P. Gate-tunable
  spin-galvanic effect in graphene-topological insulator van der Waals
  heterostructures at room temperature. \emph{Nature communications}
  \textbf{2020}, \emph{11}, 1--7\relax
\mciteBstWouldAddEndPuncttrue
\mciteSetBstMidEndSepPunct{\mcitedefaultmidpunct}
{\mcitedefaultendpunct}{\mcitedefaultseppunct}\relax
\EndOfBibitem
\bibitem[Liu \latin{et~al.}(2019)Liu, Luo, Ji, Barone, Picozzi, and
  Xiang]{bsvsp}
Liu,~K.; Luo,~W.; Ji,~J.; Barone,~P.; Picozzi,~S.; Xiang,~H. Band Splitting
  with Vanishing Spin Polarizations in Noncentrosymmetric Crystals.
  \emph{Nature Communications} \textbf{2019}, \emph{10}, 1--6\relax
\mciteBstWouldAddEndPuncttrue
\mciteSetBstMidEndSepPunct{\mcitedefaultmidpunct}
{\mcitedefaultendpunct}{\mcitedefaultseppunct}\relax
\EndOfBibitem
\bibitem[Sakano \latin{et~al.}(2020)Sakano, Hirayama, Takahashi, Akebi,
  Nakayama, Kuroda, Taguchi, Yoshikawa, Miyamoto, Okuda, Ono, Kumigashira,
  Ideue, Iwasa, Mitsuishi, Ishizaka, Shin, Miyake, Murakami, Sasagawa, and
  Kondo]{tell-prl}
Sakano,~M.; Hirayama,~M.; Takahashi,~T.; Akebi,~S.; Nakayama,~M.; Kuroda,~K.;
  Taguchi,~K.; Yoshikawa,~T.; Miyamoto,~K.; Okuda,~T.; Ono,~K.;
  Kumigashira,~H.; Ideue,~T.; Iwasa,~Y.; Mitsuishi,~N.; Ishizaka,~K.; Shin,~S.;
  Miyake,~T.; Murakami,~S.; Sasagawa,~T.; Kondo,~T. Radial Spin Texture in
  Elemental Tellurium with Chiral Crystal Structure. \emph{Phys. Rev. Lett.}
  \textbf{2020}, \emph{124}, 136404\relax
\mciteBstWouldAddEndPuncttrue
\mciteSetBstMidEndSepPunct{\mcitedefaultmidpunct}
{\mcitedefaultendpunct}{\mcitedefaultseppunct}\relax
\EndOfBibitem
\bibitem[Khokhriakov \latin{et~al.}(2018)Khokhriakov, Cummings, Song, Vila,
  Karpiak, Dankert, Roche, and Dash]{khokhriakov2018tailoring}
Khokhriakov,~D.; Cummings,~A.~W.; Song,~K.; Vila,~M.; Karpiak,~B.; Dankert,~A.;
  Roche,~S.; Dash,~S.~P. Tailoring emergent spin phenomena in Dirac material
  heterostructures. \emph{Science Advances} \textbf{2018}, \emph{4},
  eaat9349\relax
\mciteBstWouldAddEndPuncttrue
\mciteSetBstMidEndSepPunct{\mcitedefaultmidpunct}
{\mcitedefaultendpunct}{\mcitedefaultseppunct}\relax
\EndOfBibitem
\end{mcitethebibliography}


\providecommand{\latin}[1]{#1}
\makeatletter
\providecommand{\doi}
  {\begingroup\let\do\@makeother\dospecials
  \catcode`\{=1 \catcode`\}=2 \doi@aux}
\providecommand{\doi@aux}[1]{\endgroup\texttt{#1}}
\makeatother
\providecommand*\mcitethebibliography{\thebibliography}
\csname @ifundefined\endcsname{endmcitethebibliography}
  {\let\endmcitethebibliography\endthebibliography}{}
\begin{mcitethebibliography}{8}
\providecommand*\natexlab[1]{#1}
\providecommand*\mciteSetBstSublistMode[1]{}
\providecommand*\mciteSetBstMaxWidthForm[2]{}
\providecommand*\mciteBstWouldAddEndPuncttrue
  {\def\EndOfBibitem{\unskip.}}
\providecommand*\mciteBstWouldAddEndPunctfalse
  {\let\EndOfBibitem\relax}
\providecommand*\mciteSetBstMidEndSepPunct[3]{}
\providecommand*\mciteSetBstSublistLabelBeginEnd[3]{}
\providecommand*\EndOfBibitem{}
\mciteSetBstSublistMode{f}
\mciteSetBstMaxWidthForm{subitem}{(\alph{mcitesubitemcount})}
\mciteSetBstSublistLabelBeginEnd
  {\mcitemaxwidthsubitemform\space}
  {\relax}
  {\relax}

\bibitem[Zhang and Schwingenschl\"ogl(2018)Zhang, and
  Schwingenschl\"ogl]{alpha2}
Zhang,~Q.; Schwingenschl\"ogl,~U. Rashba Effect and Enriched Spin-Valley
  Coupling in $\mathrm{Ga}X$/$M{X}_{2}$ ($M$ = Mo, W; $X$ = S, Se, Te)
  Heterostructures. \emph{Phys. Rev. B} \textbf{2018}, \emph{97}, 155415\relax
\mciteBstWouldAddEndPuncttrue
\mciteSetBstMidEndSepPunct{\mcitedefaultmidpunct}
{\mcitedefaultendpunct}{\mcitedefaultseppunct}\relax
\EndOfBibitem
\bibitem[Xiang \latin{et~al.}(2019)Xiang, Ke, and Zhang]{xiang2019tunable}
Xiang,~L.; Ke,~Y.; Zhang,~Q. Tunable Giant Rashba-Type Spin Splitting in
  PtSe$_2$/MoSe$_2$ Heterostructure. \emph{Applied Physics Letters}
  \textbf{2019}, \emph{115}, 203501\relax
\mciteBstWouldAddEndPuncttrue
\mciteSetBstMidEndSepPunct{\mcitedefaultmidpunct}
{\mcitedefaultendpunct}{\mcitedefaultseppunct}\relax
\EndOfBibitem
\bibitem[Ast \latin{et~al.}(2007)Ast, Henk, Ernst, Moreschini, Falub,
  Pacil{\'e}, Bruno, Kern, and Grioni]{ast2007giant}
Ast,~C.~R.; Henk,~J.; Ernst,~A.; Moreschini,~L.; Falub,~M.~C.; Pacil{\'e},~D.;
  Bruno,~P.; Kern,~K.; Grioni,~M. Giant Spin Splitting Through Surface
  Alloying. \emph{Physical Review Letters} \textbf{2007}, \emph{98},
  186807\relax
\mciteBstWouldAddEndPuncttrue
\mciteSetBstMidEndSepPunct{\mcitedefaultmidpunct}
{\mcitedefaultendpunct}{\mcitedefaultseppunct}\relax
\EndOfBibitem
\bibitem[Ishizaka \latin{et~al.}(2011)Ishizaka, Bahramy, Murakawa, Sakano,
  Shimojima, Sonobe, Koizumi, Shin, Miyahara, Kimura, Miyamoto, Okuda,
  Namatame, Taniguchi, Arita, Nagaosa, Kobayashi, Murakami, Kumai, Kaneko,
  Onose, and Tokura]{ishizaka2011giant}
Ishizaka,~K.; Bahramy,~M.; Murakawa,~H.; Sakano,~M.; Shimojima,~T.; Sonobe,~T.;
  Koizumi,~K.; Shin,~S.; Miyahara,~H.; Kimura,~A.; Miyamoto,~K.; Okuda,~T.;
  Namatame,~H.; Taniguchi,~M.; Arita,~R.; Nagaosa,~N.; Kobayashi,~K.;
  Murakami,~Y.; Kumai,~R.; Kaneko,~Y.; Onose,~Y.; Tokura,~Y. Giant Rashba-Type
  Spin Splitting in Bulk BiTeI. \emph{Nature Materials} \textbf{2011},
  \emph{10}, 521--526\relax
\mciteBstWouldAddEndPuncttrue
\mciteSetBstMidEndSepPunct{\mcitedefaultmidpunct}
{\mcitedefaultendpunct}{\mcitedefaultseppunct}\relax
\EndOfBibitem
\bibitem[Absor \latin{et~al.}(2018)Absor, Santoso, Harsojo, Abraha, Kotaka,
  Ishii, and Saito]{absor2018strong}
Absor,~M. A.~U.; Santoso,~I.; Harsojo,; Abraha,~K.; Kotaka,~H.; Ishii,~F.;
  Saito,~M. Strong Rashba Effect in the Localized Impurity States of
  Halogen-Doped Monolayer PtSe$_2$. \emph{Physical Review B} \textbf{2018},
  \emph{97}, 205138\relax
\mciteBstWouldAddEndPuncttrue
\mciteSetBstMidEndSepPunct{\mcitedefaultmidpunct}
{\mcitedefaultendpunct}{\mcitedefaultseppunct}\relax
\EndOfBibitem
\bibitem[Liu \latin{et~al.}(2013)Liu, Guo, and Freeman]{liu2013tunable}
Liu,~Q.; Guo,~Y.; Freeman,~A.~J. Tunable Rashba Effect in Two-Dimensional
  LaOBiS$_2$ Films: Ultrathin Candidates For Spin Field Effect Transistors.
  \emph{Nano Letters} \textbf{2013}, \emph{13}, 5264--5270\relax
\mciteBstWouldAddEndPuncttrue
\mciteSetBstMidEndSepPunct{\mcitedefaultmidpunct}
{\mcitedefaultendpunct}{\mcitedefaultseppunct}\relax
\EndOfBibitem
\bibitem[Li \latin{et~al.}(2017)Li, Wei, Zhao, Huang, and
  Dai]{li2017electronic}
Li,~F.; Wei,~W.; Zhao,~P.; Huang,~B.; Dai,~Y. Electronic and Optical Properties
  of Pristine and Vertical and Lateral Heterostructures of Janus MoSSe and
  WSSe. \emph{Journal of Physical Chemistry Letters} \textbf{2017}, \emph{8},
  5959--5965\relax
\mciteBstWouldAddEndPuncttrue
\mciteSetBstMidEndSepPunct{\mcitedefaultmidpunct}
{\mcitedefaultendpunct}{\mcitedefaultseppunct}\relax
\EndOfBibitem
\end{mcitethebibliography}

\begin{suppinfo}
The Supporting Information contains (1) Different lateral stacking configurations of Bi$_2$Se$_3$/PtSe$_2$ vdW heterostructure considered to obtain minimum-energy configuration. (2) Layer-projected band structure of Bi$_2$Se$_3$/PtSe$_2$ vdW heterostructures to show type-II and type-III band alignments. (3) Band structures of few-layer Bi$_2$Se$_3$. (4) Band structures of supercells of Bi$_2$Se$_3$ and PtSe$_2$. (5) Atomic orbital-projected band structures of Bi$_2$Se$_3$/PtSe$_2$ vdW heterostructures to show atomic-hybridization coming from different atoms. (6) Table for the Rashba spin-splittings parameters. 
\end{suppinfo}

\end{document}


\beginsupplement

\begin{figure}[!htb]
\includegraphics[width=0.9\textwidth]{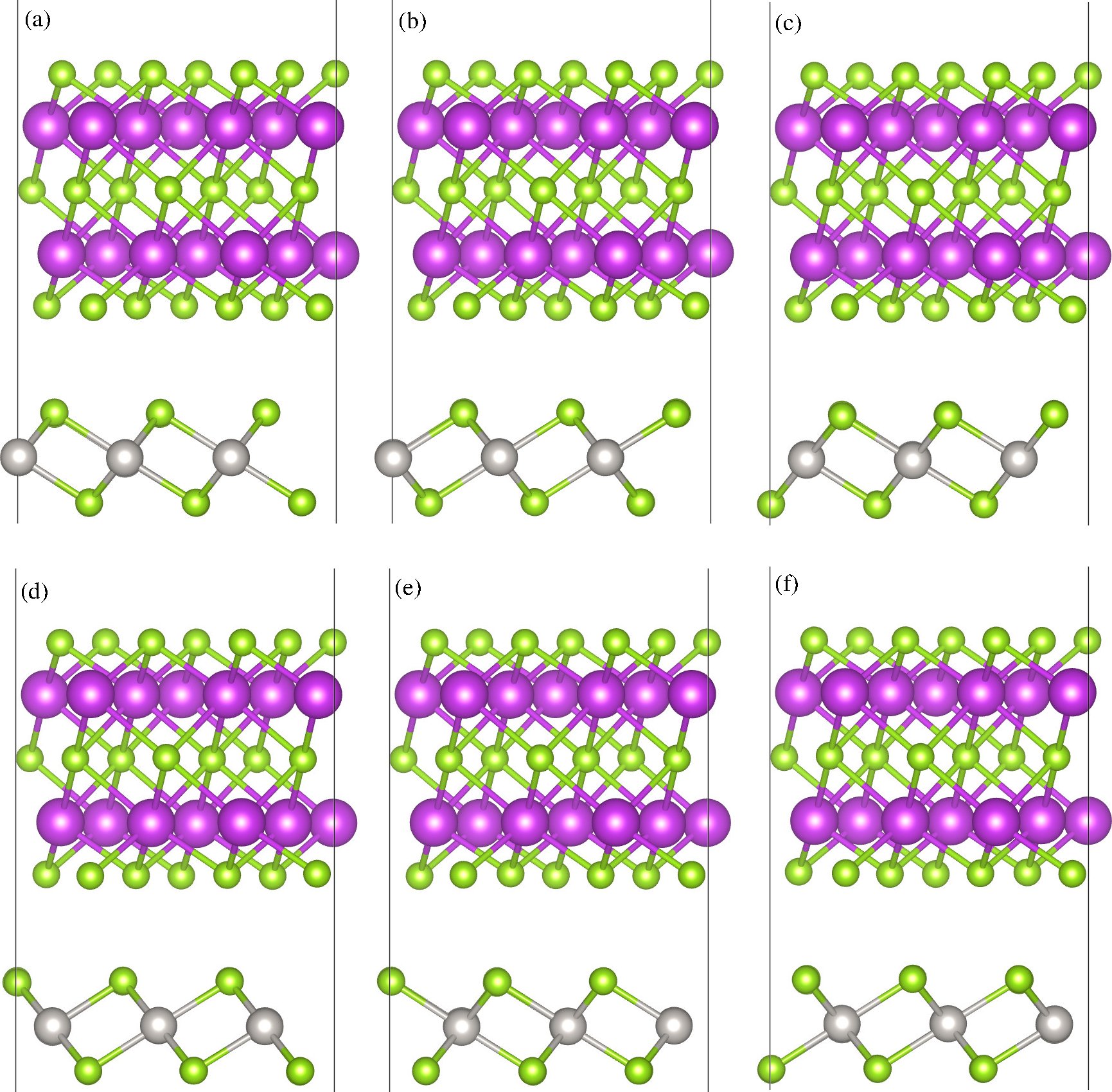}
\caption{(a-f) Different lateral stacking configurations considered to arrive at the minimum energy configuration. The total energies fall in the range of 0 to 10 meV for these stackings.}
\label{fig:figs1}
\end{figure}

\textbf{\begin{figure}[!htb]
\includegraphics[width=0.9\textwidth]{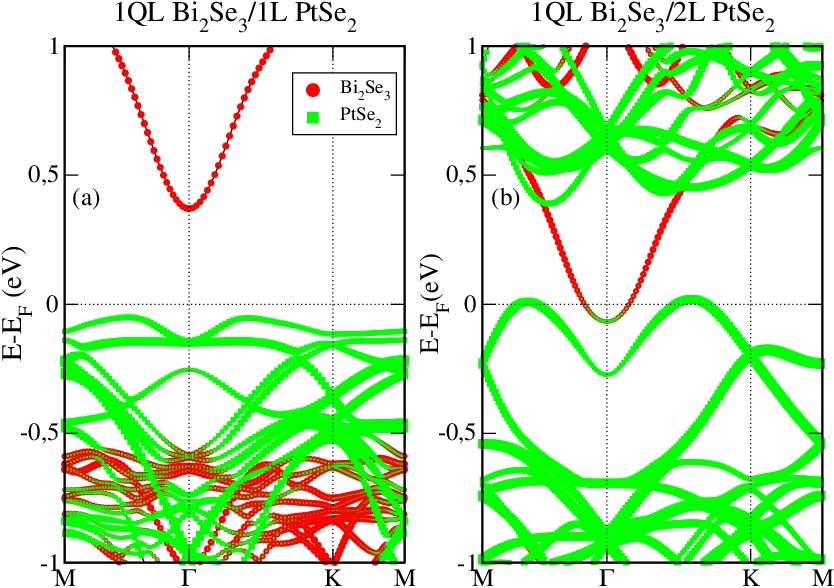}
\caption{Layer-projected band structures showing type-II and type-II band alignments in (a) 1QL Bi$_2$Se$_3$/1L PtSe$_2$ and (b) 1QL Bi$_2$Se$_3$/2L PtSe$_2$ vdw heterostructures, respectively. }
\label{fig:figs2}
\end{figure}}

\begin{figure}[!htb]
\includegraphics[width=0.9\textwidth]{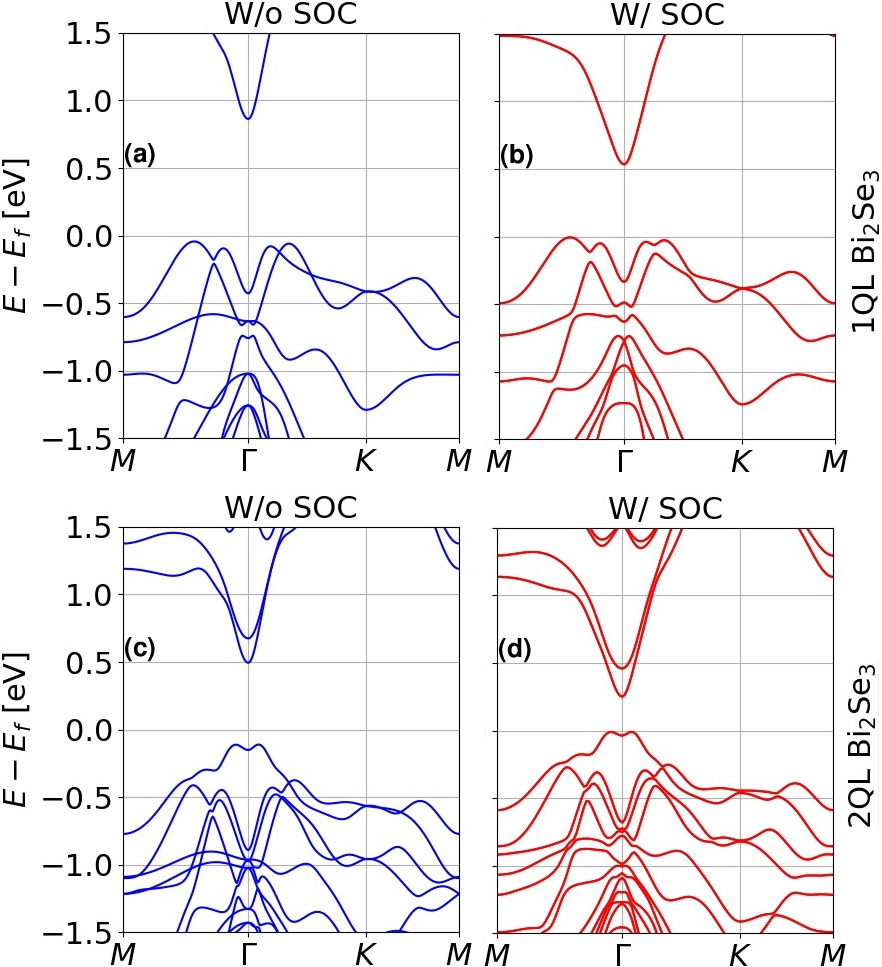}
\caption{Band structures of (a,b) 1QL Bi$_2$Se$_3$, and (c,d) 2QL Bi$_2$Se$_3$ (without SOC (left,blue) and with SOC (right,red)).}
\label{fig:figs3}
\end{figure}

\begin{figure}[!htb]
\includegraphics[width=0.9\textwidth]{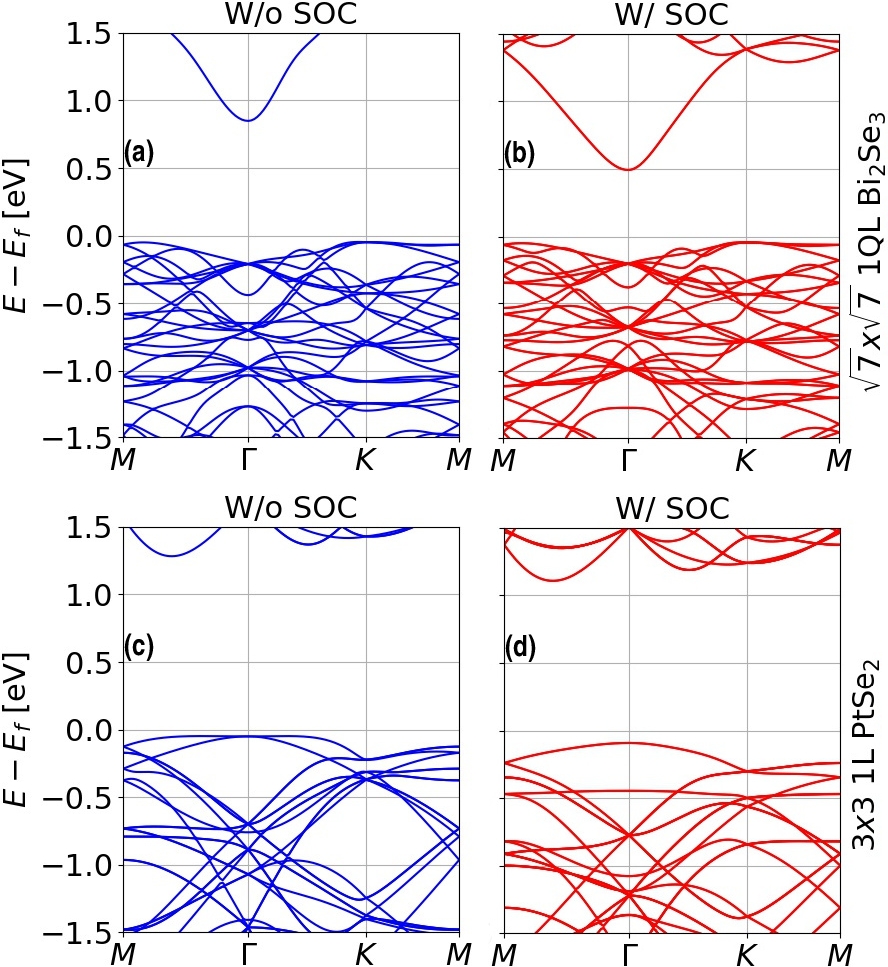}
\caption{Band structures of (a,b) $\sqrt{7}\times\sqrt{7}$ 1QL Bi$_2$Se$_3$, and (c,d) $3\times3$ 1L PtSe$_2$ (without SOC (left,blue) and with SOC (right,red)).}
\label{fig:figs4}
\end{figure}

\begin{figure}[!htb]
\includegraphics[width=0.9\textwidth]{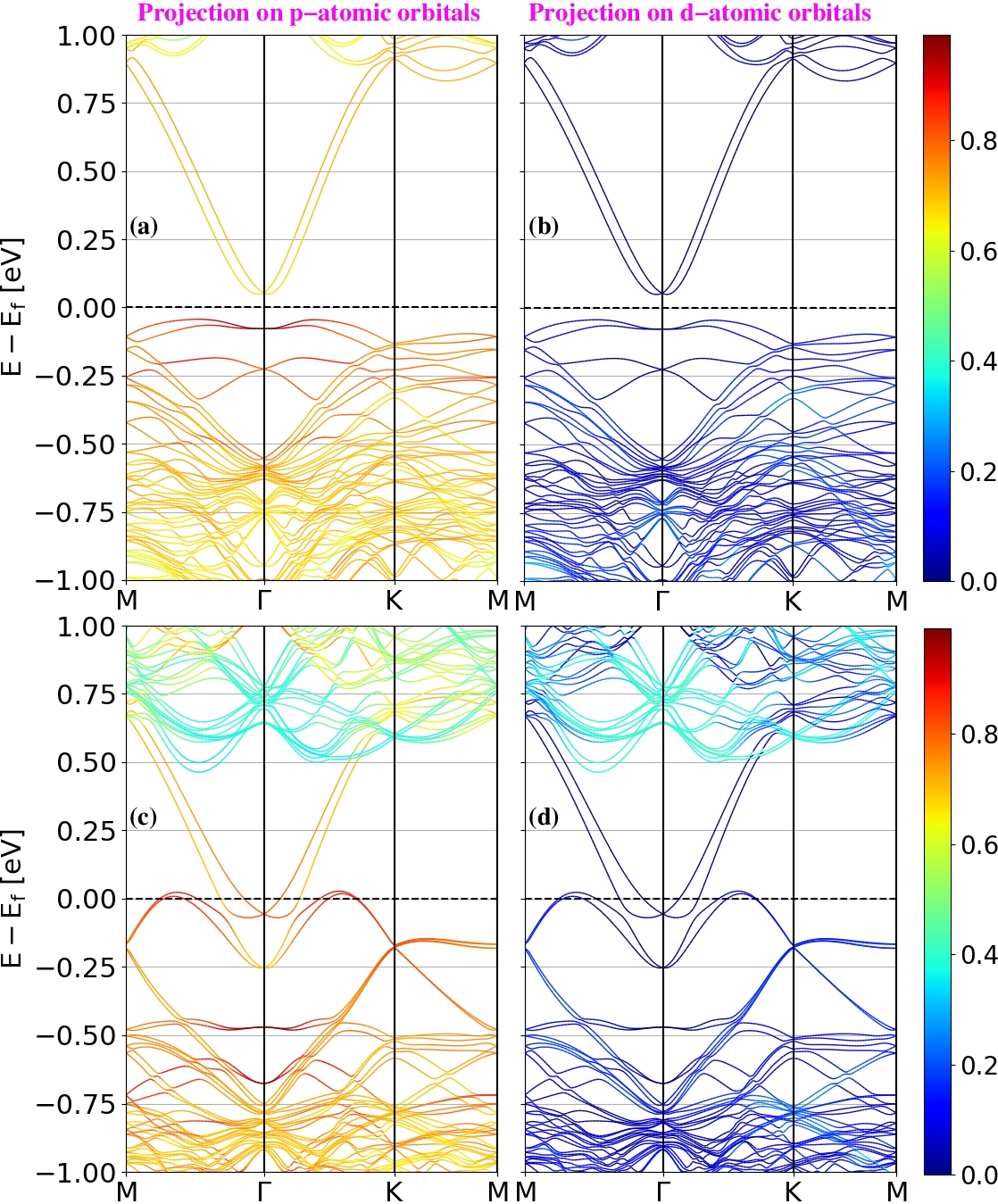}
\caption{Atomic orbital-projected band structures of (a,b) 1QL Bi$_2$Se$_3$/1L PtSe$_2$ and (c,d) 1QL Bi$_2$Se$_3$/2L PtSe$_2$ vdw heterostructures, respectively. SOC is incorporated in all cases.}
\label{fig:figs2prime}
\end{figure}

\begin{table}[!ht]
\caption{Rashba spin-splitting parameters of few-layer Bi$_2$Se$_3$/PtSe$_2$ vdW heterostructures in comparison to previous works. $E_R$ is energy difference between the CB/VB minimum/maximum and band crossing at the $\Gamma$-point, $k_0$ is the momentum shift and $\alpha_R$ is the Rashba parameter.} 
\centering 
\begin{tabular}{c  c  c  c  c} 
\hline\hline 
Case & $E_R(\text{meV})$ & $k_0(\mathrm{\AA}^{-1})$ & $\alpha_R(\text{eV}\,\mathrm{\AA})$ & Reference \\ [0.5ex] 
\hline 
1QL Bi$_2$Se$_3$/1L PtSe$_2$       & 4.8 & 0.002 & 4.80  (CB) &This work\\
2QL Bi$_2$Se$_3$/1L PtSe$_2$       & 4.0 & 0.002 & 4.00  (CB) &This work\\
2QL Bi$_2$Se$_3$/1L PtSe$_2$       & 15  & 0.006 & 5.00  (CB) &This work\\ 
                $-$                & 275 & 0.033 & 16.66 (VB) &This work\\
2QL Bi$_2$Se$_3$/2L PtSe$_2$       & 11  & 0.005 & 4.40  (CB) &This work\\
                $-$                & 278 & 0.033 & 16.84 (VB) &This work\\ [1ex]
GaSe/MoSe$_2$ vdW heterostructure  & 31  & 0.13 & 0.49  (VB) & \cite{alpha2}\\
PtSe$_2$/MoSe$_2$ vdW heterostructure & 150 & 0.23 & 1.30  (VB)& \cite{xiang2019tunable}\\
Bi/Ag(111) surface alloy              & 200 & 0.13 & 3.05  (VB)& \cite{ast2007giant}\\
Bulk BiTeI                            & 100 & 0.052& 3.80 (CB) & \cite{ishizaka2011giant}\\
I-doped PtSe$_2$                      & 12.5 & 0.015 & 1.70  (CB)& \cite{absor2018strong}\\
LaOBiS$_2$                            & 38   & 0.025 & 3.04 (VB) & \cite{liu2013tunable}\\
MoSSe                                 & 1.4  & 0.005 & 0.53  (VB)& \cite{li2017electronic}\\[1ex] 
\hline 
\end{tabular}
\label{table:tables1} 
\end{table}

\clearpage
\bibliography{supp.bib}